# Dynamical Evolution of the Didymos–Dimorphos Binary Asteroid as Rubble Piles following the DART Impact

Harrison F. Agrusa[1], Fabio Ferrari[2], Yun Zhang[3,4], Derek C. Richardson[1], and Patrick Michel[4]
[1] Department of Astronomy, University of Maryland, College Park, MD 20742, USA; hagrusa@astro.umd.edu
[2] Space Research and Planetary Sciences, Physics Institute, University of Bern, Bern, 3012, Switzerland
[3] Department of Aerospace Engineering, University of Maryland, College Park, MD 20742, USA
[4] Université Côte dAzur, Observatoire de la Côte dAzur, CNRS, Laboratoire Lagrange, Nice, France

## Abstract

Previous efforts have modeled the Didymos system as two irregularly shaped rigid bodies, although it is likely that one or both components are in fact rubble piles. Here, we relax the rigid-body assumption to quantify how this affects the spin and orbital dynamics of the system following the DART impact. Given the known fundamental differences between our simulation codes, we find that faster rigid-body simulations produce nearly the same result as rubble-pile models in scenarios with a moderate value for the momentum enhancement factor, $\beta$ ($\beta \sim 3$) and an ellipsoidal secondary. This indicates that the rigid-body approach is likely adequate for propagating the post-impact dynamics necessary to meet the DART Mission requirements. Although, if Dimorphos has a highly irregular shape or structure, or if $\beta$ is unexpectedly large, then rubble-pile effects may become important. If Dimorphos's orbit and spin state are sufficiently excited, then surface particle motion is also possible. However, these simulations are limited in their resolution and range of material parameters, so they serve as a demonstration of principle, and future work is required to fully understand the likelihood and magnitude of surface motion.

*Unified Astronomy Thesaurus concepts:* Asteroid dynamics (2210); Asteroid rotation (2211); Asteroid satellites (2207); Natural satellite dynamics (2212); Orbits (1184); Near-Earth objects (1092); N-body simulations (1083); N-body problem (1082); Celestial mechanics (211)

## 1. Introduction

Launched on 2021 November 24, and arriving at the Didymos system on 2022 September 26, NASA's Double Asteroid Redirection Test (DART) will be the first planetary defense-oriented mission to conduct a kinetic impact on an asteroid (Cheng et al. 2018; Rivkin et al. 2021). The target of the mission is Dimorphos, the smaller component of the binary asteroid 65803 Didymos. The DART spacecraft will intercept Dimorphos, making an approximate head-on impact, which will slow Dimorphos's relative speed, leading to a reduction in the mutual semimajor axis and orbit period of the binary system. The change in orbit period can then be measured using ground-based observations (Naidu et al. 2022; Pravec et al. 2022) to ultimately infer the momentum enhancement factor, $\beta$, a unitless quantity that describes how much net momentum is transferred to the target body (Rivkin et al. 2021; Stickle et al. 2022). Prior to DART's kinetic impact, the spacecraft will deploy the Light Italian Cubesat for Imaging of Asteroids (LICIAcube) cubesat, which will fly by the system, imaging the early stages of the cratering process as well as provide images of the system from different angles to improve Dimorphos's shape determination, and possibly provide information on the surface characteristics (Dotto et al. 2021; Cheng et al. 2022; Pajola et al. 2022). Lead by the European Space Agency, the Hera spacecraft will then visit Dimorphos 4 yr after the DART impact. Hera consists of an orbiter and two cubesats, called Juventas and Milani, that will fully characterize the physical (including interior), compositional, and dynamical states of the system (Michel et al. 2018). Hera will also further assess the impact effects, in particular the size and morphology of the crater left by DART in addition to a precise measurement of Dimorphos's mass in order to improve the estimate of $\beta$ (Michel et al. 2022). Together, DART and Hera constitute the Asteroid Impact and Deflection Assessment cooperation between the two space agencies.

The shape, spin, and size of Didymos in addition to the mutual orbit period and average separation of the binary are relatively well understood (Scheirich & Pravec 2009; Fang & Margot 2012; Naidu et al. 2020; Scheirich & Pravec 2022). However, both the shape and spin state of Dimorphos are poorly constrained. It is often assumed that Dimorphos is in a relaxed, tidally locked state, meaning that its spin period matches the measured mutual orbit period of ∼11.9216 hr, and its libration amplitude is small, although this has not yet been directly confirmed (Richardson et al. 2022). Regardless of Dimorphos's preimpact dynamical state, DART's perturbation to the mutual orbit will likely excite Dimorphos's spin state as a result of the high degree of spin–orbit coupling due to the irregular shapes of both components and their close proximity. The degree of excitation will be highly dependent on the magnitude of $\beta$, in addition to Dimorphos's shape, which is assumed to be a triaxial ellipsoid and commonly parameterized by its ellipsoidal axis ratios $a/b$ and $b/c$, with $a \geqslant b \geqslant c$. Using rigid, full two-body simulations, Agrusa et al. (2021) demonstrated that Dimorphos can become attitude unstable, and its spin state could evolve chaotically as a result of the impact, depending on its shape (i.e., moments of inertia). However, it is unclear how relaxing the rigid-body formalism to allow the bodies to behave as rubble piles will affect the binary dynamics and attitude-stability properties of Dimorphos.

The fast rotation rate and oblate shape of Didymos is indicative of a rubble-pile structure (Walsh 2018). Furthermore, if Dimorphos has a common origin (YORP-induced fission or mass loss and subsequent reaccumulation in orbit, for example), then it is quite plausible that both bodies are rubble piles (Walsh et al. 2008; Jacobson & Scheeres 2011). Therefore, extending previous rigid-body studies to include granular physics while maintaining high-fidelity modeling of the mutual spin and orbital dynamics represents an important next step in simulating the post-impact evolution of the Didymos binary. This work presents our first steps at studying and constraining the dynamical implications of a rubble-pile treatment. Therefore, we focused our efforts on understanding the limits at which rubble-pile effects are important. Assuming that Didymos and/or Dimorphos are confirmed to be rubble piles upon DART's arrival, the parameter space of possible body shapes, particle-size distributions, etc., will also be greatly reduced, allowing us to eventually have a better handle on the relative importance of rubble-pile effects.

As a point of clarification, this work focuses on the dynamical evolution of the system when one or both bodies are treated as rubble piles and allowed to deform *over time*. In this work, we do not consider immediate deformation due to the DART impact itself. We refer the reader to the companion papers by Hirabayashi et al. (2022) and Nakano et al. (2022) that model the direct deformation of Didymos or Dimorphos due to the DART impact and propagate the resulting system as rigid bodies. The degree of shape deformation that DART will cause is unclear, as it depends on many unknowns such as the bulk density, cohesion, boulder distribution, among many other parameters. Recent numerical simulations by Raducan & Jutzi (2022) indicate that a DART-scale impact could significantly deform and resurface Dimorphos if it has low cohesive strength. This nonzero possibility of significant impact-induced shape change is something to consider in future work, depending on the outcome of the DART impact. Combined imagery from DART and LICIAcube will be crucial in determining the impact outcome prior to Hera's arrival.

In Section 2, we introduce the simulation codes employed in this work and briefly describe how these simulations are set up. Section 3 compares the mutual dynamics of the system when treated as rubble piles versus rigid bodies and finds that rigid-body and rubble-pile models are in broad agreement in typical circumstances. Then, Section 4 explores the limits at which the rubble-pile structure of Didymos or Dimorphos may affect the dynamics and, conversely, the limits at which the dynamics may affect the structure of either body. Finally, Section 5 summarizes this research and discusses future work to follow after the DART impact.

## 2. Methodology

We study the dynamics of the Didymos binary system with self-gravitating, rubble-pile models of Didymos and Dimorphos using the *N*-body granular physics codes PKDGRAV and GRAINS. In addition, the GUBAS full two-body problem (F2BP) code is used as an additional point of comparison with rigid-body results. For convenience, we provide the current best estimates for the physical and dynamical parameters at the time of this writing in Table 1. These parameters can also be found in Appendix A of Rivkin et al. (2021), but are updated here with the latest values as of this writing based on the Design Reference Asteroid v. 3.2 (DRA; DART mission internal document). These values may not necessarily match *exactly* those used in the simulations presented here, although they are all close and within the estimated uncertainties of the system. The three codes used in this work are briefly described below.

**Table 1**
Selected Dynamical Parameters from Rivkin et al. (2021) and Updated with the Latest Values of This Writing (DRA v. 3.2)

| Parameter | Value |
| --- | --- |
| Volume-equivalent Diameter of Primary $D_P$ | $780 \pm 30$ m |
| Volume-equivalent Diameter of Secondary $D_S$ | $164 \pm 18$ m |
| Bulk Densities of Components $\rho_P$ | $2170 \pm 350$ kg m$^{-3}$ |
| Mean Separation of Component Centers $a_{\rm orb}$ | $1.20 \pm 0.03$ km |
| Secondary Shape Elongation $a/b$, $b/c$ | $1.3 \pm 0.2$, 1.2 (assumed) |
| Total Mass of System $M$ | $(5.55 \pm 0.42) \times 10^{11}$ kg |
| Secondary Orbital Period $P_{\rm orb}$ | $11.921\,628\,9 \pm 0.0000028$ h |
| Secondary Orbital Eccentricity $e_{\rm orb}$ | $<0.03$ |
| Primary Rotation Period $P_P$ | $2.260\,0 \pm 0.0001$ h |
| Secondary Rotation Period $P_S$ | $P_{\rm orb}$ (assumed tidally locked) |
| Secondary Orbital Inclination $i_{\rm orb}$ | $0°$ (assumed) |

### 2.1. GUBAS

The General Use Binary Asteroid Simulator (GUBAS) is an open-source[5] simulation tool that can quickly solve for the coupled spin and orbital motion of two arbitrarily shaped rigid masses with high fidelity (Davis & Scheeres 2020, 2021). GUBAS has been benchmarked against other F2BP codes, formally adopted for rigid-body modeling of the Didymos system for the DART mission (Agrusa et al. 2020), and successfully used to model the Didymos–Dimorphos binary (Agrusa et al. 2021; Meyer et al. 2021). In this work, we consider the motion of the two bodies solely under their mutual gravity.

### 2.2. PKDGRAV

PKDGRAV is a massively parallel *N*-body tree code that can represent each component of the Didymos system as an aggregate of many spherical particles (Richardson et al. 2000; Stadel 2001). In this work, the *k*-d tree code is not used, so the gravitational forces on each particle are computed by summing directly over all particles at each time step to ensure the highest possible accuracy at the expense of computational speed (i.e., the full $\mathcal{O}(N^2)$ *N*-body problem). The contact forces on interacting particles are handled using the soft-sphere discrete element method (SSDEM), which allows for particles to slightly overlap each other with a mediating spring force as a proxy for particle deformation (Schwartz et al. 2012). With SSDEM, a user can set parameters such as the restoring spring constant and coefficients of rolling and twisting friction to achieve the desired material properties (Zhang et al. 2017). Following common practice, we set the spring constant such that the overlap between two particles never exceeds 1% of the smallest particle's radius. As will be discussed later, this approach may lead to an artificially deformable body, but it allows us to adequately resolve interparticle contacts without having to resort to prohibitively short time steps. It is also possible to include interparticle cohesive forces in PKDGRAV (Zhang et al. 2018). Here, we ignore any potential cohesion in Dimorphos in order to observe the maximum possible effect of its rubble-pile treatment. In all PKDGRAV simulations presented herein, we select the friction parameters that represent a gravel-like material and yield a friction angle of ∼38° (Zhang et al. 2018). PKDGRAV has already been used and validated in rigid F2BP studies of the Didymos system (Agrusa et al. 2020), and here we extend our analysis by enabling the code's SSDEM feature to fully model Didymos and Dimorphos as rubble piles, rather than rigid bodies.

---
[5] The code is available at https://github.com/alex-b-davis/gubas and can be effectively run with a single core on a desktop computer.

### 2.3. GRAINS

GRAINS is a N-body code that accounts for both gravitational and granular physics interactions between a large number of nonspherical particles (Ferrari et al. 2017). Gravity computations are done using either a direct $N^2$ algorithm, which accounts for the mutual gravity between all particles in the system, or by using a GPU-based octree implementation of the Barnes–Hut algorithm (Ferrari et al. 2020). As done with PKDGRAV, we do not use the tree implementation in order to ensure the highest accuracy of mutual gravity computations in this work. Contact and collision interactions are handled using the Smooth Contact Method module of the open-source multiphysics code CHRONO (Tasora et al. 2016). GRAINS has been used recently to study the stability and internal structure of Didymos (Ferrari & Tanga 2022; Hirabayashi et al. 2022). In this work, we use GRAINS to model Dimorphos as a gravitational aggregate of irregularly shaped, meter-sized boulders. Each boulder has a different polyhedral shape with about 10 vertices on average, and an aspect ratio (smallest to largest dimension) between 0.7 and 1. The individual shape of each fragment is built as the convex hull of a randomly generated cloud of points. This makes each boulder unique, although all of them are similar in size. The contact parameters are set to reproduce the properties of gravel-like material and are based on previous benchmarking studies (Fleischmann et al. 2015; Pazouki et al. 2017; Ferrari & Tanga 2020). The cohesion between fragments is set to zero in GRAINS, as cohesive effects are not considered in this work.

### 2.4. Problem Setup

Due to the high computational cost of N-body problems, the number of cases and integration times had to be limited. Due to this constraint, we employ several simulation approaches to model different aspects of the system's dynamics. All things being equal, PKDGRAV simulations run much faster than GRAINS due to PKDGRAV's treatment of constituent particles as spheres rather than polyhedral shapes. Therefore we rely on PKDGRAV to conduct long-term simulations (∼1 yr) or for simulating both bodies as rubble piles at high resolution. In this work, GRAINS is employed in specific shorter-term cases to understand the influence of irregular particle shapes where computationally feasible. In general, the long-term simulations that focus on Dimorphos's spin and attitude properties use PKDGRAV and treat Didymos as a point mass to increase the computational speed. The exact details and initial conditions for each set of simulations are explained within their respective subsections to avoid confusion.

In all simulations presented herein, we assume that Dimorphos's preimpact spin period matches the observed mutual orbit period. Although a synchronously rotating Dimorphos has not been directly confirmed with radar and photometry, there is good theoretical and observational evidence to indicate that this is the most likely preimpact state for Dimorphos. This assumption is addressed with more detail in a companion paper by Richardson et al. (2022). Some of the theoretical justifications include the relatively low frequency of natural impacts and close planetary flybys (Fuentes-Muñoz & Scheeres 2020; Meyer & Scheeres 2021), and efficient tidal dissipation in rubble piles (Goldreich & Sari 2009; Nimmo & Matsuyama 2019). The observational evidence includes a measured upper limit on the binary eccentricity of 0.03 (Scheirich & Pravec 2022), as well as the abundance of synchronous rotators in other close binary systems (Pravec et al. 2016). Furthermore, the best-fit orbital solution indicates a quadratic drift in the mean anomaly of $\Delta M_d = 0.15 \pm 0.14$ deg yr$^{-2}$(3σ), implying that outward tidal expansion is being overcome by inward BYORP drift (Scheirich & Pravec 2022). If BYORP is acting in the system, then this would require a tidally locked secondary (Ćuk & Burns 2005). However, there are also good theoretical arguments that a nonsynchronous spin state could be easily excited and long lived (Ćuk et al. 2021; Quillen et al. 2022). If, upon DART's arrival, there is a reasonable indication that Dimorphos is in nonsynchronous rotation, then new models will be needed to incorporate this effect.

Furthermore, in all simulations, we assume that the DART impact will impart an instantaneous $\Delta v$ to Dimorphos's orbital velocity. We neglect any instantaneous changes to Dimorphos's spin that could result from an off-center impact that imparts a torque to the body, as such a torque is expected to be relatively small. If, however, DART impacts farther than expected from the body's center-of-mass, then accounting for this torque may be a topic of future work, depending on the impact outcome. Further discussion of this assumption, as well as other approximations for the dynamics of the system, is available in Richardson et al. (2022).

### 3. Spin and Orbital Dynamics as Rubble Piles

As previously mentioned, it is not feasible to treat both bodies as rubble piles over long integrations while maintaining high resolution and numerical accuracy. Therefore, we take two approaches to model the system, i.e., the *full-rubble-pile* approach, in which both Didymos and Dimorphos are modeled as rubble piles, and the *single-rubble-pile* approach, where only Dimorphos is a rubble pile and Didymos is treated as a point mass. The full-rubble-pile approach is applied to assess the dynamics on short timescales (days to weeks), and the single-rubble-pile approach can be used to study the system on longer timescales (years). Although the latter approach fails to capture any dynamical effects due to Didymos's higher-order gravity moments, it still adequately captures the attitude-stability properties of Dimorphos, since this is predominantly determined by its own shape and the mutual eccentricity (Agrusa et al. 2021; Wisdom 1987). However, the higher-order gravity effects, such as those due to Didymos's $J_2$ and $C_{22}$ gravity coefficients are captured with the "full-rubble-pile" approach as well as rigid-body simulations presented in other works (Agrusa et al. 2021; Meyer et al. 2021; Richardson et al. 2022).

#### 3.1. Long-term Dynamics with a Point-mass Didymos

Agrusa et al. (2021) found that the attitude stability of Dimorphos is highly sensitive to its shape (i.e., moments of

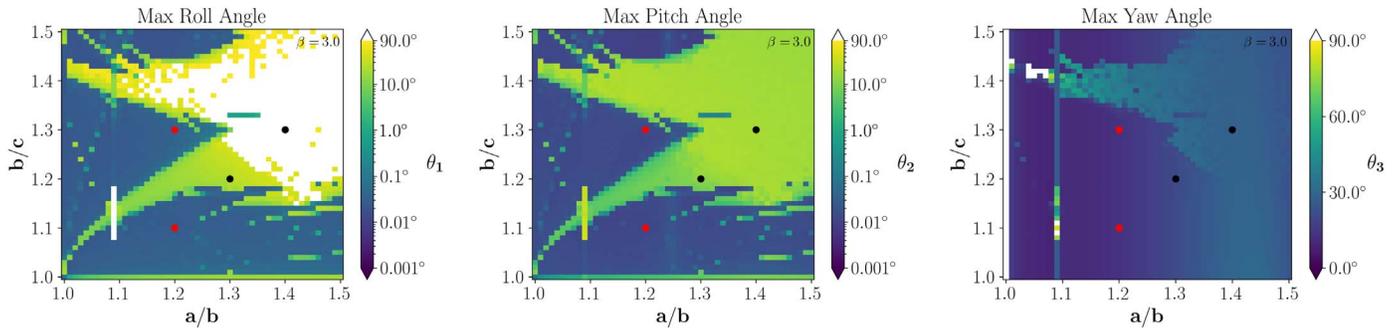

**Figure 1.** From Agrusa et al. (2021), the maximum Euler angles (roll, pitch, and yaw) achieved by Dimorphos over a 1 yr simulation in GUBAS following a DART-like perturbation to the mutual orbit with $\beta = 3$. The overlaid dots show the four shapes under consideration in this study with the two expected attitude-stable cases (`ab1.2bc1.3` and `ab1.2bc1.1`) in red and the two expected attitude-unstable cases (`ab1.4bc1.3` and `ab1.3bc1.2`) in black. The structure of the attitude-stability region is a result of various resonances between the mean motion and Dimorphos's libration, spin precession, and nutation frequencies.

**Table 2**
The Body Shapes for Dimorphos Considered in This Section, Informed by the Larger Parameter Sweep Used in Agrusa et al. (2021)

| Name | $a/b$ | $b/c$ | $a$ [m] | $b$ [m] | $c$ [m] | $N$ | $M_B$ [$10^9$ kg] | Agrusa et al. (2021) Prediction for $\beta = 3$ |
|---|---|---|---|---|---|---|---|---|
| `ab1.4bc1.3` | 1.40 | 1.30 | 111.32 | 79.51 | 61.16 | 3737 | 4.878 80 | Unstable |
| `ab1.3bc1.2` | 1.30 | 1.20 | 103.16 | 79.35 | 66.13 | 3739 | 4.883 06 | Unstable |
| `ab1.2bc1.3` | 1.20 | 1.30 | 100.44 | 83.70 | 64.39 | 3742 | 4.884 70 | Stable |
| `ab1.2bc1.1` | 1.20 | 1.10 | 95.00 | 79.17 | 71.97 | 3719 | 4.887 22 | Stable |

**Note.** $a/b$ and $b/c$ are the ellipsoidal axis ratios for each body shape along with their physical lengths, $a$, $b$, and $c$. We also report the total number of particles, $N$, that make up the body, as well as the total mass of the body, $M_B$. The body masses vary slightly in the simulation in order to produce a dynamically relaxed preimpact state of the system. Note that all four bodies are constructed from the same power-law particle-size distribution and that the bulk size of Dimorphos remains conserved between all cases.

inertia). This is a result of resonances that can occur among the orbital, libration, spin precession, and nutation frequencies of Dimorphos that can trigger attitude instabilities. Aside from the orbital frequency (mean motion), these frequencies depend directly on Dimorphos's moments of inertia (i.e., shape), which are unknown. In order to extend the results of Agrusa et al. (2021), we select four possible ellipsoidal shapes for Dimorphos: two in which previous rigid-body simulations predict the presence of an attitude instability, and two of which are expected to remain stable at an eccentricity of $e \sim 0.025$ (i.e., a perturbation to a circular orbit consistent with $\beta = 3$). The four selections for the shape of Dimorphos are listed on Table 2. These four cases were selected in order to sample both the unstable and stable configurations in many parts of the parameter space surveyed by Agrusa et al. (2021). Given that nonprincipal axis rotation of Dimorphos is commonly observed in simulations (Agrusa et al. 2021; Ćuk et al. 2021; Quillen et al. 2022), comparing cases across different attitude-stability regimes is important for benchmarking the rigid-body and rubble-pile approaches. Figure 1 shows a modified version of a plot from Agrusa et al. (2021) that displays the maximum Euler angles achieved by Dimorphos for a year following a DART-like impact in which $\beta = 3$ based on GUBAS rigid-body simulations. Overlaid on the plot are the four Dimorphos shapes considered in this section. Renderings of the four shapes are shown on Figure 2. Case `ab1.4bc1.3` is the most elongated shape considered in this study, and is expected to enter the "barrel instability," a phenomenon in which a satellite enters a rolling state about its long axis, which is described in greater detail in Ćuk et al. (2021) and indicated by the roll angle exceeding 90° in Figure 1. Case `ab1.3bc1.2` was selected to provide an additional attitude-unstable case and because it is the nominal shape used by the mission in the DRA. Due to the lack of observational constraints for Dimorphos, this particular shape was selected for the DRA based on the observed elongations of the secondary component of other binary systems (Pravec et al. 2016). Finally, cases `ab1.2bc1.3` and `ab1.2bc1.1` were selected due to their strong attitude-stability properties according to GUBAS rigid-body simulations and to see whether the rubble-pile codes reproduce the same predicted stability.

Since the attitude-stability properties of Dimorphos are driven primarily by its own shape, accounting for the oblate shape of Didymos is not necessary, allowing us to simply treat it as a point mass. The set of PKDGRAV runs presented in this subsection performs a given simulation two times: once as a rigid body in which all the constituent particles of Dimorphos are locked together into a rigid aggregate; and once using PKDGRAV's SSDEM package where Dimorphos is treated as a rubble pile. This allows for a direct *apples-to-apples* comparison in which the rigid and SSDEM cases have identical initial shapes. We also simulate a matching case in GUBAS, although these cases do not match *exactly* because Dimorphos's shape is treated as an idealized ellipsoid. When interpreting the results of this section, the difference between a GUBAS and a rigid PKDGRAV case shows the fundamental differences between the two codes and their respective shape representation of Dimorphos, while the differences between the SSDEM and rigid PKDGRAV cases show any effects due to the deformability of Dimorphos.

This leads to one additional complication: when the rubble-pile Dimorphos is placed into orbit around Didymos, it suddenly feels a tidal force that acts to slightly deform its shape into a new equilibrium configuration. This change is extremely small, but in order to directly compare with the SSDEM and rigid PKDGRAV cases, we must make sure both bodies have the *exact same* starting shape. This is achieved by simulating the preimpact system for 24 hr (∼2 orbit periods) with Dimorphos as a rubble pile, which is sufficiently long for the particles to reach a new relaxed state. Then the simulation is halted, the constituent particles of Dimorphos are locked into a rigid aggregate (or left alone for the SSDEM cases), and a $\Delta v$ is applied to the orbital velocity of Dimorphos as a proxy for the DART impact. Finally, the simulation is restarted and allowed to propagate for 1 yr of simulation time.

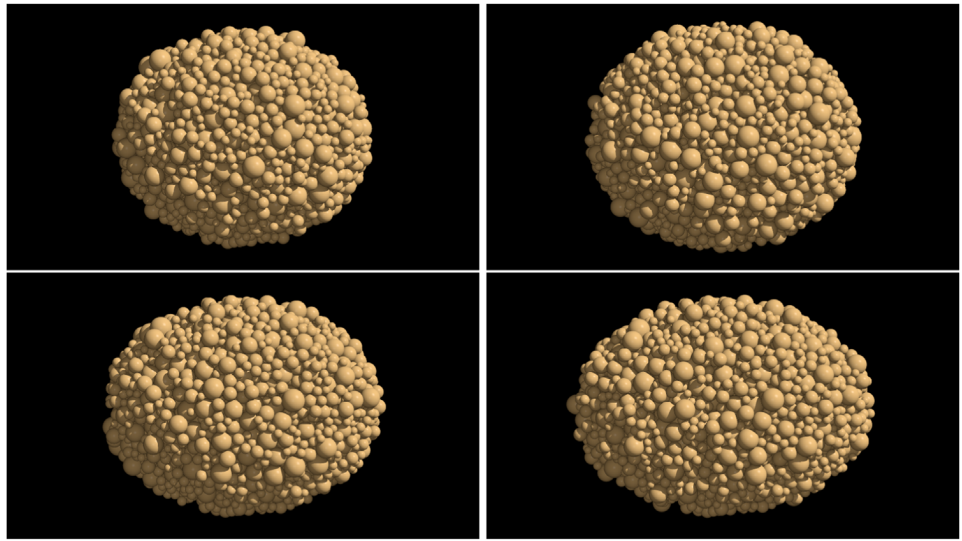

**Figure 2.** The four ellipsoidal representations of Dimorphos in PKDGRAV considered in this section. Each body is drawn from the same power-law particle-size distribution with an index of −3 and an average particle radius of ∼4.2 m. These are *top-down* views, meaning that the camera is situated above Dimorphos's spin pole. Top left: `ab1.2bc1.1`. Top right: `ab1.2bc1.3`. Bottom left: `ab1.3bc1.2`. Bottom right: `ab1.4bc1.3`.

To summarize, simulations are set up as follows:

1. PKDGRAV particles are generated from a power-law particle-size distribution with an index of −3.0 and an average particle radius of 4.2 m. The maximum and minimum possible particle radii are 8.4 and 2.8 m. The power-law index was chosen to be similar to the boulder-size distribution exponents observed on the surfaces of several asteroids (e.g., Michikami et al. 2010; Walsh et al. 2019; Michikami et al. 2019), and the size cutoffs were chosen such that each realization of Dimorphos would contain ∼4,000 particles to keep the computational costs reasonable. The cloud of particles is allowed to collapse (with all friction parameters set to zero) to form an approximately spherical body. Then the desired ellipsoidal shape for Dimorphos is *carved out* of this rubble pile by simply deleting any particles that lie outside of its surface.

2. After Dimorphos is carved into its desired shape, friction is turned on, and the body is simulated on its own (no Didymos) to allow it to come to an equilibrium configuration under self-gravity and spin.

3. The initial conditions for the binary orbit were generated using GUBAS and the optimization routine described in Agrusa et al. (2021). In this case, the optimization routine used a point-mass primary and ellipsoidal secondary to derive the initial conditions such that the resulting system has a spin-synchronous secondary with a mutual semimajor axis and orbital period that match their respective observed values in Table 1.

4. The simulation is then started in PKDGRAV and stopped after 24 hr of simulation time (∼2 orbit periods). This allows for the rubble-pile model of Dimorphos to come to equilibrium after it suddenly feels the tidal stress resulting from being placed in an orbit around the point-mass Dimorphos.

5. After all settling is complete, Dimorphos is given an instantaneous $\Delta v$ as a proxy for the DART impact. The change in velocity is determined based on the expected mass and relative velocity of the DART spacecraft, as well as a guess for $\beta$ ($\beta = 3$ is assumed for these particular simulations).[6] Here, we assume that DART imparts all of its momentum within the mutual orbit plane and opposite Dimorphos's motion. Due to DART's expected near-head-on impact geometry, this planar approximation is not expected to significantly alter the dynamics (Richardson et al. 2022).[7]

6. The simulation is then restarted and run for 1 yr of integration time, once as a rigid body and once using the SSDEM feature. These simulations are run with a time step of ∼0.86 s, which is small enough to adequately resolve interparticle contact interactions and is also well below the 1.875 s time step that was found to be appropriate for accurately modeling the mutual orbital dynamics with PKDGRAV by Agrusa et al. (2020).

---

[6] The mass and relative velocity of the DART spacecraft are assumed to be 535 kg and 6.6 km s$^{-1}$, respectively. Since the time this investigation was begun, the best estimate of the mass and relative velocity has changed to 536 kg and 6.15 km s$^{-1}$ respectively. Therefore, for a given value of $\beta$ and mass of Dimorphos, these simulations slightly overestimate the $\Delta v$ that Dimorphos will receive.

[7] For a simplified, head-on impact, Dimorphos receives $\Delta v = -\beta \frac{M_{\text{DART}} v_{\text{DART}}}{M_B}$, where $M_{\text{DART}}$ and $v_{\text{DART}}$ are the spacecraft's respective mass and velocity, and $M_B$ is Dimorphos's mass.

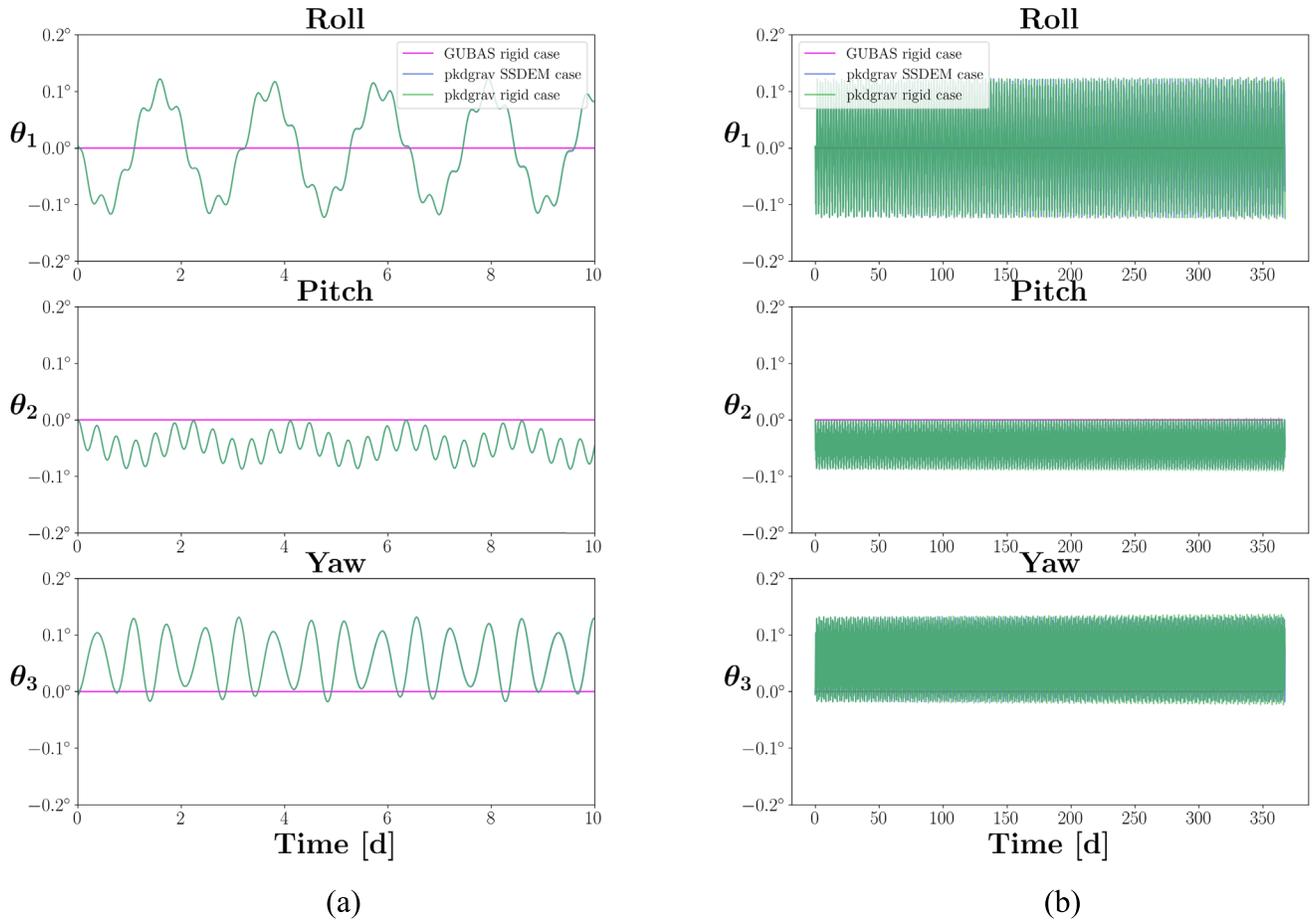

**Figure 3.** Evolution of Dimorphos's Euler angles for `ab1.2bc1.1` when the system is in a dynamically relaxed state. The GUBAS rigid case shows *perfect* tidal locking since the system is idealized: a circular orbit, spherical primary, and smooth ellipsoidal secondary. The PKDGRAV simulations (both as a rigid aggregate and SSDEM body) are not quite as relaxed since their mass distributions do not perfectly match the equivalent GUBAS simulations.

### 3.1.1. Attitude-stable Cases

Based on rigid-body simulations, we expect cases `ab1.2bc1.1` and `ab1.2bc1.3` to remain attitude stable against a DART-like impact. Here we show that PKDGRAV simulations reproduce the same behavior, both as a rigid body and as a rubble pile using SSDEM.

Figure 3 shows Dimorphos's Euler angles (roll, pitch, and yaw) when it has the shape `ab1.2bc1.1` and the system is in the relaxed state (i.e., no DART impact). The GUBAS results show *perfect* tidal locking (all Euler angles being zero) due to the idealized symmetry: a spherical primary, a smooth ellipsoidal secondary, and a circular orbit. As discussed earlier, the rubble-pile realization of Dimorphos does not *exactly* match the mass distribution of the GUBAS case, meaning that the dynamics are not perfectly relaxed. This is also the reason that the pitch and yaw angles oscillate around a nonzero value. Although the Euler angle amplitudes are nonzero, the PKDGRAV simulations are in qualitative agreement with GUBAS, with any minor deviations being attributable to differences in the shape representations. Furthermore, the rigid and SSDEM PKDGRAV results are nearly identical and hard to distinguish, meaning that the rigid-body approximation is more than adequate in this scenario, as we would expect.

We also show the same model for Dimorphos (`ab1.2bc1.1`) following a DART-like perturbation consistent with $\beta = 3$ on Figure 4. The DART perturbation is applied at $t = 1$ day and leads to an eccentricity of the mutual orbit of $e \sim 0.025$. The effect of this can be seen on the yaw-angle plot in Figure 4(a), where Dimorphos starts librating after 1 day. Both PKDGRAV simulations as well as GUBAS show attitude stability and the same libration amplitude (given by yaw angle). Furthermore, the rigid and SSDEM PKDGRAV cases are nearly identical, again indicating that Dimorphos is behaving as a rigid body and that the rigid-body approximation is adequate for simulating the post-impact attitude dynamics in this case.

Figure 5 shows the post-impact spin evolution of Dimorphos for $\beta = 3$ but with the `ab1.2bc1.3` shape representation. Here, we see similar behavior: the PKDGRAV cases show strong qualitative agreement with the matching GUBAS simulation, with minor deviations due to differences in the body shape representations and the codes themselves. Furthermore, the rigid and SSDEM PKDGRAV cases show nearly identical behavior, indicating that the rigid-body approximation is valid in the regime of attitude stability.

### 3.1.2. Attitude-unstable Cases

Previous work has shown that Dimorphos's spin state can evolve chaotically after its attitude becomes unstable. This means that Dimorphos's spin evolution is highly sensitive to any small changes in the system or the initial conditions. Therefore, it is impossible for the PKDGRAV and GUBAS simulations to match exactly since they do not have the *exact* same initial conditions, nor the same numerical integrator. However, we find *qualitative* agreement between the GUBAS simulation and both the rigid and SSDEM PKDGRAV simulations for `ab1.3bc1.2` and `ab1.4bc1.3`. Generally speaking, if GUBAS predicts a given configuration to be attitude unstable, we see the same behavior with PKDGRAV.

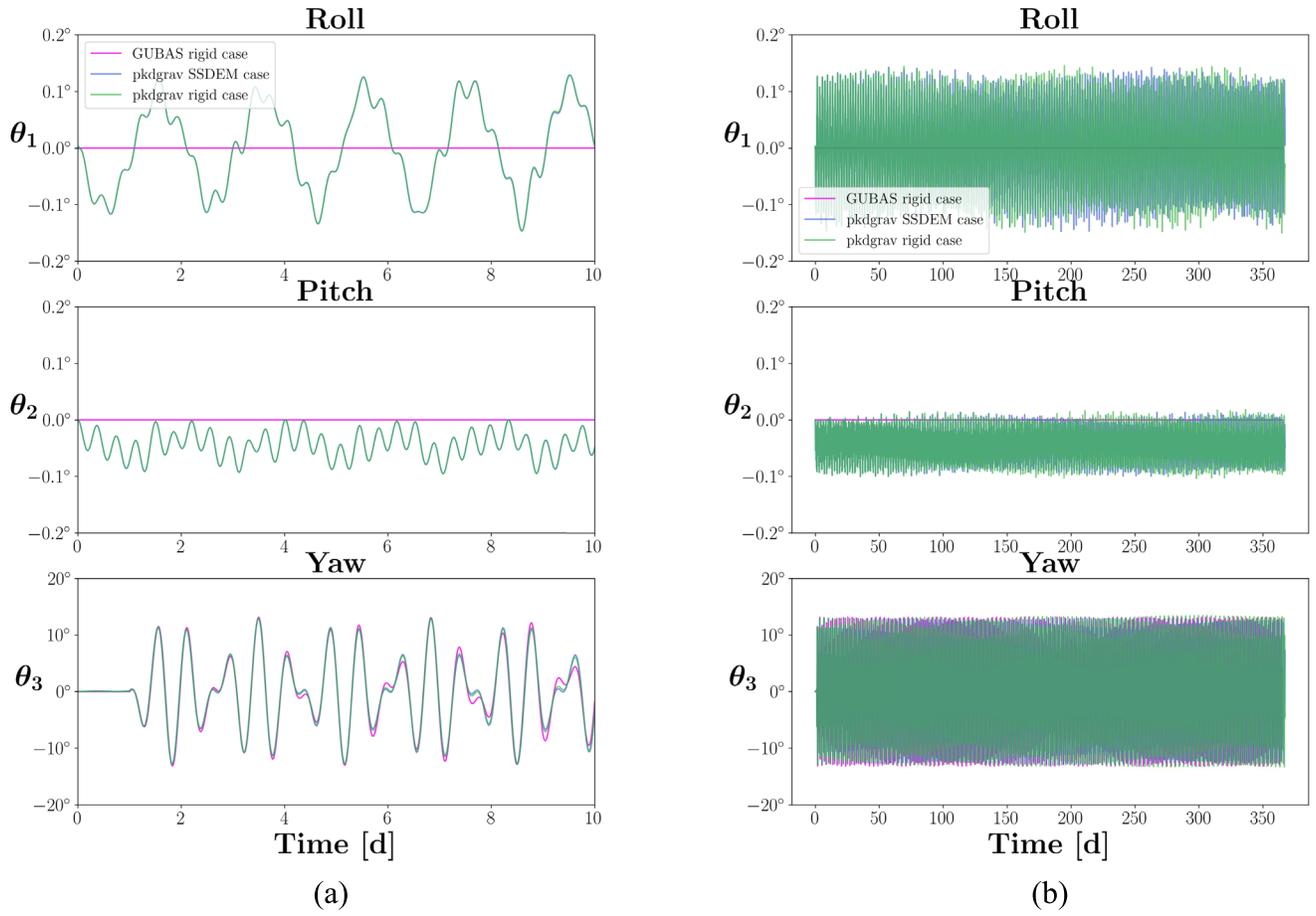

**Figure 4.** Evolution of Dimorphos's Euler angles for `ab1.2bc1.1` when Dimorphos's orbital velocity is perturbed consistent with a DART impact with $\beta = 3$. In other words, the mutual orbit has an eccentricity of ∼0.025. The DART impact is applied at $t = 1$ days, which is when Dimorphos begins significantly librating.

Figure 6 shows the post-impact evolution of Dimorphos's Euler angles with the shape `ab1.3bc1.2` after a DART-like perturbation of $\beta = 3$. Both PKDGRAV simulations and GUBAS reveal that Dimorphos should become attitude unstable for this shape and show approximately the same post-impact libration amplitude of ∼20° (yaw angle). In addition, they are in broad agreement for the amplitude of roll and pitch oscillations once Dimorphos enters the attitude instability. The main difference between the codes is in the timing of the instability, which is technically impossible to predict given the chaotic nature of the system. Additionally, the codes use different numerical integrators, simulated on different machines, and the rubble-pile Dimorphos does not have the *exact* same mass distribution as the idealized ellipsoid, so there is no reason for us to expect the timing of the instability to match.

Following a DART-like impact, `ab1.4bc1.3` is expected to not only be attitude unstable but also enter the so-called "barrel instability," characterized by a rotation about the secondary's long axis (Ćuk et al. 2021). Indeed, PKDGRAV finds the same behavior both when Dimorphos is treated as a rigid body and a deformable rubble pile, as seen in Figure 7. Although each case enters the instability at different times, all three cases show the same qualitative behavior, with Dimorphos episodically rotating about its long axis (roll angle hitting 180°) and the pitch and yaw amplitudes capped at ∼20° and ∼25°, respectively.

In conclusion, the rubble-pile approach reproduces the same qualitative behavior seen in equivalent rigid-body simulations. This indicates that the rigid-body approach is appropriate, at least for moderate values of $\beta$ and an ellipsoidal secondary.

### 3.2. Short-term Dynamics with Full Rubble Piles

Increasing in complexity, we then simulate the system with *both* Didymos and Dimorphos modeled as rubble piles. A rubble-pile model composed of 13,049 particles is constructed based on the radar-derived shape model of Didymos (Naidu et al. 2020) to capture its irregular top-like shape. This choice of particle number represents the optimal compromise between the model resolution and computational cost. The sizes of these particles range from ∼7.8 to ∼31.2 m following a differential power-law distribution with an exponent of −3 (the same as for Dimorphos in the single-rubble-pile runs). Dimorphos is assumed to be a rubble-pile ellipsoid consisting of 504 particles with the same power-law distribution but smaller sizes, i.e., ∼5.4 to ∼16.0 m, to better characterize its shape.[8] Given that the tidal interaction raised by the primary is expected to drive the secondary's spin toward synchronization much more efficiently than in the case of a monolithic secondary and that the presence of a rubble-pile primary can cause changes in the orbital semimajor axis and eccentricity more rapidly (Goldreich & Sari 2009), we focus on a post-impact attitude-unstable case in this section to investigate the effect of the rubble-pile binary. A shape with $a \approx 109.93$ m, $b \approx 81.41$ m, $c \approx 60.21$ m, is selected to represent the shape of Dimorphos (similar to the attitude-unstable `ab1.4bc1.3` case shown in Table 2).

---
[8] Due to the increased computational cost of simulating Didymos as a rubble pile, we simulate Dimorphos at lower particle resolution than in the previous section.

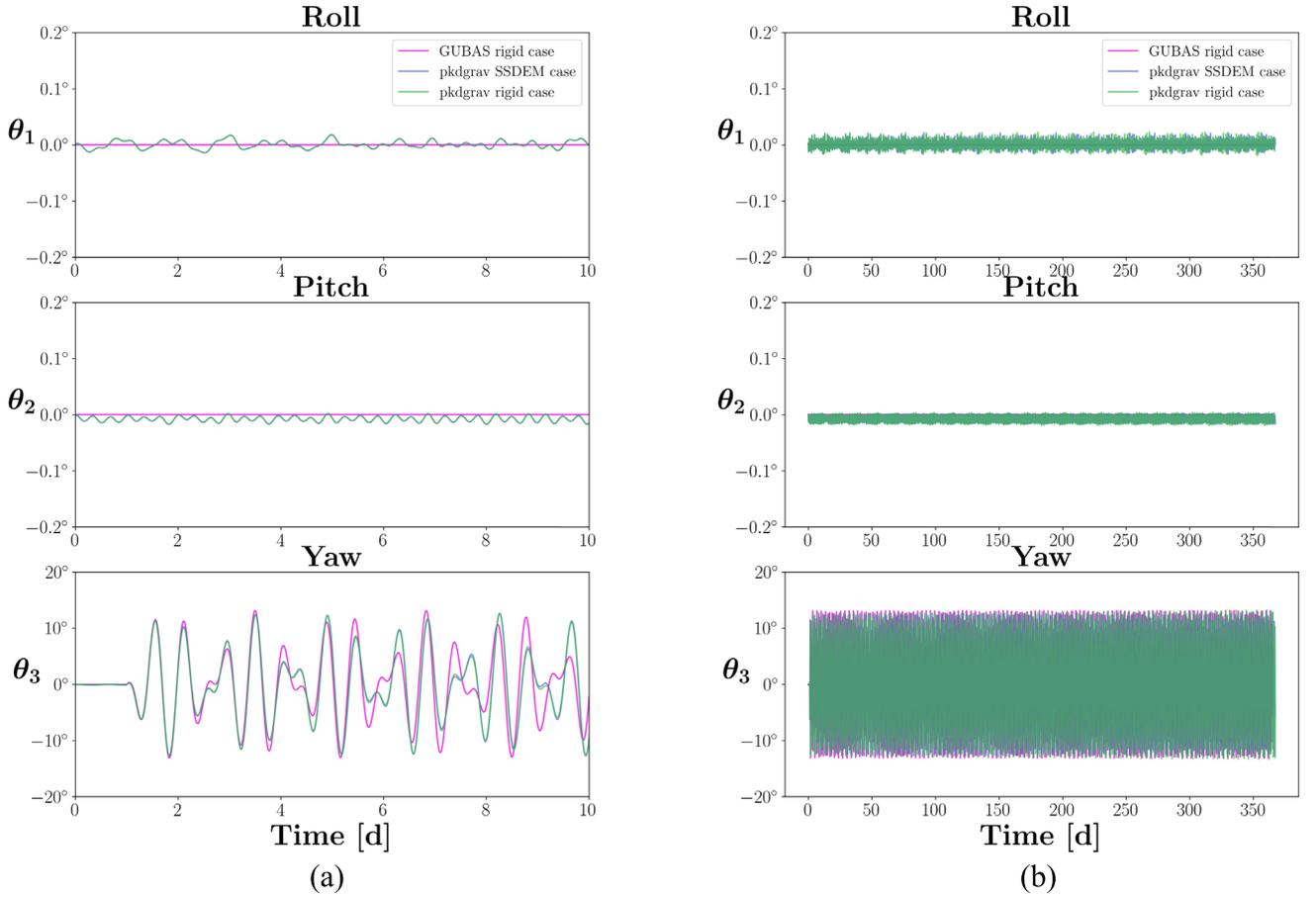

**Figure 5.** Evolution of Dimorphos's Euler angles for `ab1.2bc1.3` when Dimorphos's orbital velocity is perturbed consistent with a DART impact with $\beta = 3$. In other words, the mutual orbit has an eccentricity of ∼0.025. The DART impact is applied at $t = 1$ day, which is when Dimorphos begins significantly librating.

The two rubble-pile models were generated via the gravitational collapse and shape-carving procedure introduced in Section 3.1. A quasi-static spin-up procedure (see Section 2.2 in Zhang et al. 2021) was applied to settle the two models separately to their respective equilibrium state with the corresponding spin rate, i.e., ∼2.26 hr for Didymos and ∼11.91 hr for Dimorphos. Then, the two models were combined in one simulation by assigning the movement of their mass centers according to the orbital dynamics of the Didymos–Dimorphos system derived from the point-mass-Didymos approach (see Section 3.2). Due to its fast rotation, the Didymos rubble pile needs some amount of material cohesion to maintain its stability at its assumed density. Adopting from our previous study (Zhang et al. 2021), we use a macroscopic cohesion of 20 Pa and a friction angle of 38° to represent the material properties of Didymos. Dimorphos is modeled with the same friction angle but zero cohesion to provide for the maximum possible effect of the rubble-pile structure. The full-rubble-pile models were allowed 24 hr to settle down under their mutual gravity, and then the velocity of Dimorphos was modified along the instantaneous orbital direction to capture the momentum-change effect of the DART impact (the same procedure as introduced in Section 3.1). Figure 8 shows the initial configuration of the full rubble-pile model. Simulations were run for 30 days to reveal the short-term dynamics of the rubble-pile structure.

To investigate the effect of the impact momentum transfer efficiency, we carried out simulations for one case without the DART impact and three cases with the DART impact and different values of $\beta$. Figure 9 shows the evolution of the spin periods, axis ratios, coordination number, and various orbital parameters of Dimorphos. Considering that the binary orbit deviates from a Keplerian orbit, the orbital period is evaluated as the time it takes for Dimorphos to complete each successive 360° revolution, following the orbit period formalism of Meyer et al. (2021). The orbit is time-varying due to periodic exchanges of angular momentum between the mutual orbit and Dimorphos's spin.

As shown in Figure 9, without being perturbed by the DART impact, the state of Dimorphos stays near the nominal observational values oscillating with small magnitudes due to the deviation of Didymos's gravity field from the point mass, as shown by the blue curves. The rubble-pile Dimorphos behaves like an elastic body, and its shape ($c/a$) and averaged contact number ($N_c$) expand and shrink in response to its orbital position and rotational state. The orbit and rotation can be further excited by the DART impact. As shown by the curves marked with different $\beta$ values, the oscillation magnitudes after DART impact increase substantially and monotonically with larger $\beta$. In the case of $\beta = 1$, the averaged orbital period of Dimorphos decreases to $\sim 11.7$ hr, meaning that the DART impact causes a $\sim 720$ s change, which is one order of magnitude larger than the Level 1 requirement of the DART mission Rivkin et al. (2021). With $\beta = 3$, this change increases to $\sim 1800$ s, and the center separation between the two bodies is decreased by $\sim 0.03$ km on average. The orbital eccentricity is elevated immediately due to the instantaneous change of orbital speed and then oscillates as angular momentum is exchanged between the mutual orbit and Dimorphos's spin state. Due to the irregular gravity field, the orbital evolution of Dimorphos is not perfectly planar, and the orbital inclination can also be excited by the DART impact, despite an assumed planar impact. These results are consistent with previous analyses based on rigid-body dynamics (Agrusa et al. 2021; Meyer et al. 2021; Richardson et al. 2022).

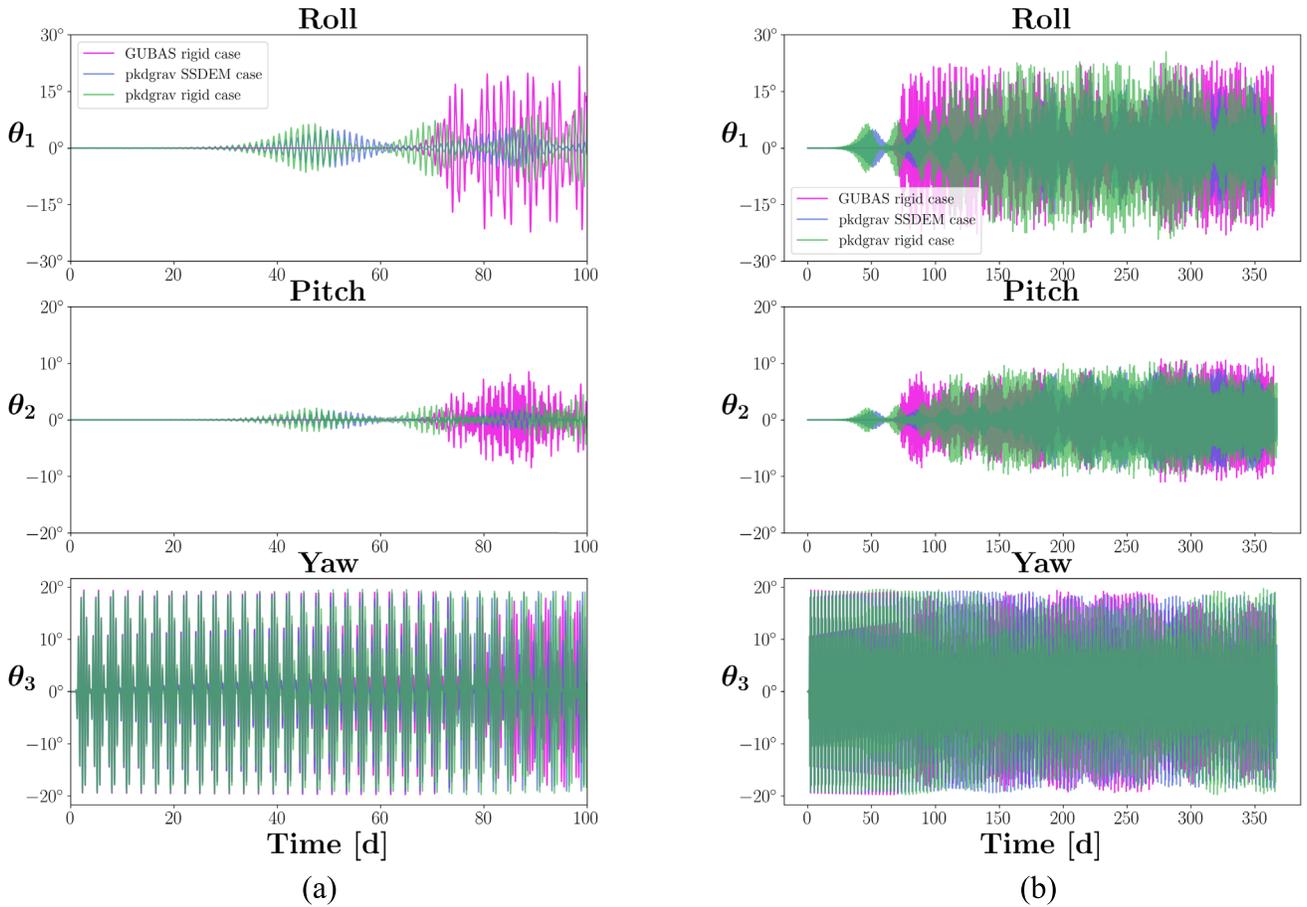

**Figure 6.** Evolution of Dimorphos's Euler angles for `ab1.3bc1.2` for $\beta = 3$. The GUBAS and PKDGRAV simulations are in broad agreement, both showing attitude instability with about the same amplitudes in the three Euler angles. The deviations between the three simulations are attributable to differences between the codes and the shape representation of Dimorphos.

To better understand the role that the full-rubble-pile model plays in the binary system dynamics, we carried out matching rigid-body simulations using GUBAS with both Didymos and Dimorphos modeled as dynamically equivalent equal-volume ellipsoids (DEEVEs). The body masses, DEEVE shapes, positions, velocities, spins, and orientations of the GUBAS model are set to match the initial state of the full-two-rubble-pile model. Figures 10 and 11 compare the post-impact spin and orbital evolution of Dimorphos for the unperturbed and three $\beta$ cases predicted by the full-two-rubble-pile model against the GUBAS rigid-body model. In general, these two models show strong qualitative agreement with each other (even quantitative in the case of $\beta = 1, 2$; the small deviation in the inclination is mainly due to the different treatment in approximating the non-point-mass gravity field of Didymos). The consistency in the eccentricity evolution indicate that the variations in $e_O$ mainly come from the exchange of angular momentum between Dimorphos's rotational and orbital states, rather than tidal dissipation due to their rubble-pile treatment. This agrees with the previous theoretical understanding of the rubble-pile–tidal interaction, whose timescales on affecting the binary dynamics would be on the order of megayears (Goldreich & Sari 2009). The barrel instability predicted by the rigid-body approach is also observed in the rubble-pile simulation, although Dimorphos never completes a full rotation about its long axis over the 30 days time frame. Given that the barrel instability is observed in the 1 yr simulations presented in Section 3.1, this is likely due to the chaotic nature of its spin evolution rather than the fact that we are considering the full-rubble-pile approach here. These results further confirm that the rigid-body approximation is valid in terms of predicting the general attitude evolution and stability of the binary system.

## 4. Limits of the Rigid-body Approach

In Section 3, we showed that under moderate conditions (a typical $\beta$ value and ellipsoidal-shaped secondary) the rubble-pile models are in broad, qualitative agreement with equivalent rigid-body simulations. This likely means that faster rigid F2BP codes are adequate for predicting the dynamical state of the system following the DART impact. In this section, we explore the limit at which this assumption may break down and rubble-pile models of the system are required. DART's arrival at the Didymos system will confirm whether the two components are in fact rubble piles. If they are rubble piles (as we expect), the DART imagery will also greatly reduce the range of possible shapes for Dimorphos as well as constrain its boulder-size–frequency distribution, which will provide a much better picture of the importance of rubble-pile effects. Therefore, the results presented in this section are just preliminary and meant to guide future studies of the binary system following the DART impact.

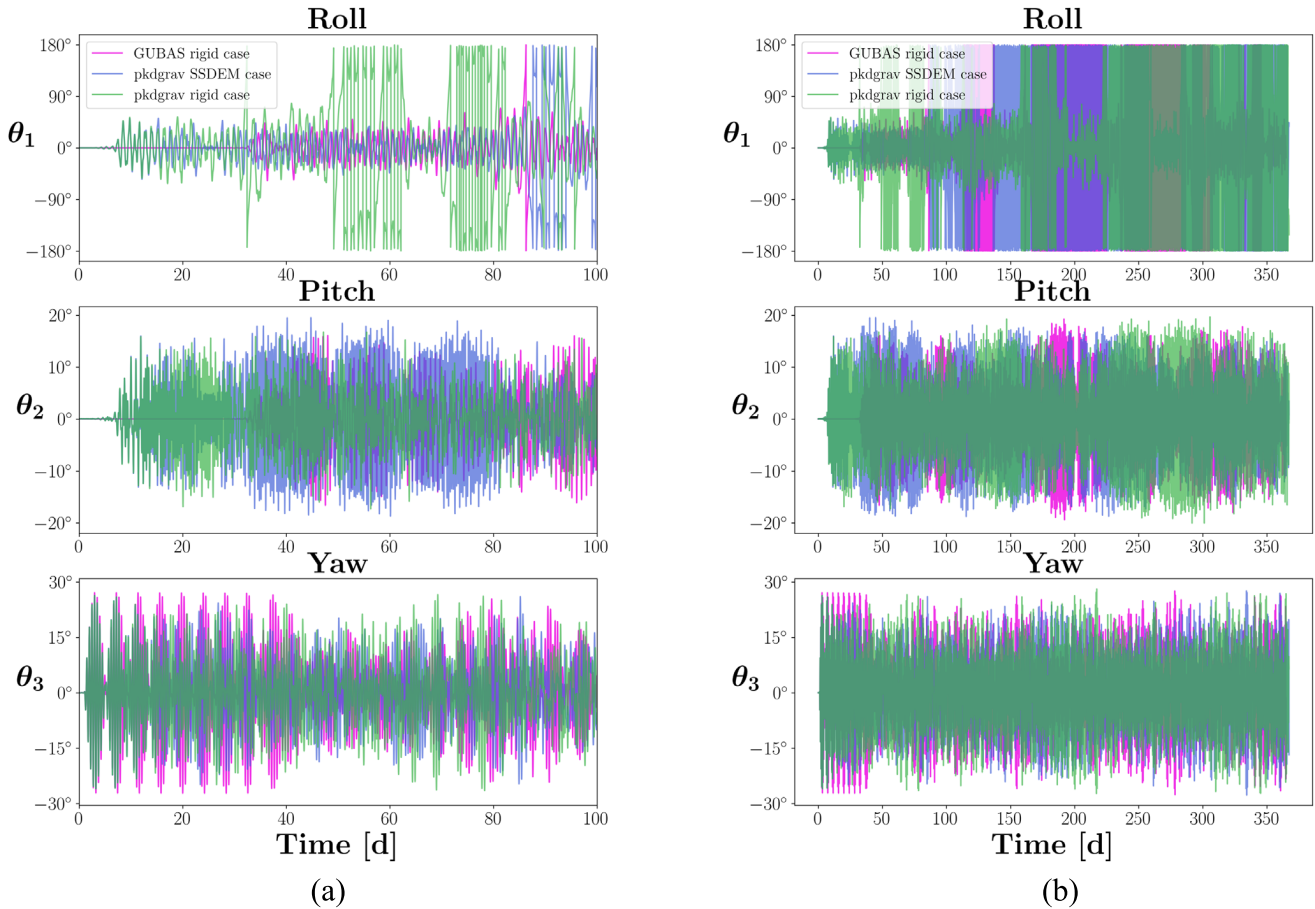

**Figure 7.** Evolution of Dimorphos's Euler angles for `ab1.4bc1.3` for $\beta = 3$. The GUBAS and PKDGRAV simulations are in broad agreement, both showing the barrel instability, although the instability occurs at different times. The deviations between the three simulations are attributable to differences between the codes and the shape representation of Dimorphos.

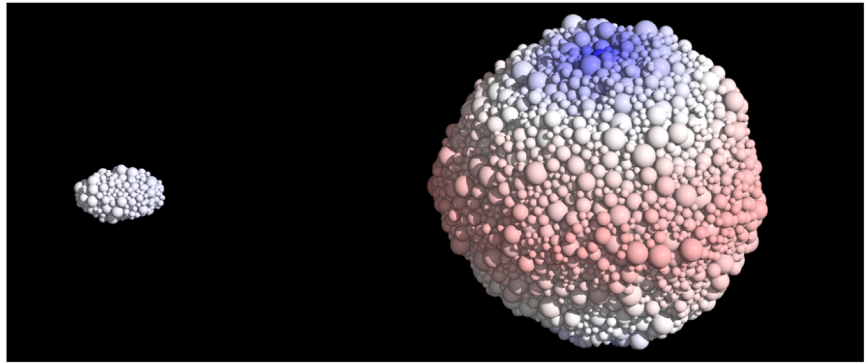

**Figure 8.** Rubble-pile representations of Didymos and Dimorphos in PKDGRAV considered in the full-rubble-pile approach (Section 3.2). Each body is drawn from the same power-law particle-size distribution with an index of −3 and particle radii of ∼7.8 to ∼31.2 m (Didymos) and ∼5.4 to ∼16.0 m (Dimorphos). Particles are color-coded by the speed magnitude to show the spin orientation of Didymos and the relative orbital speed of Dimorphos. The camera is tilted ∼23° above the equatorial plane to better present the two bodies's shapes.

### 4.1. Limits on Strong Tidal Effects

As a result of DART's near-head-on impact, the orbital speed of Dimorphos will be reduced, causing it to fall on a tighter orbit with a decreased pericenter distance. In this context, it is worth exploring the tidal interactions between Dimorphos and Didymos, to investigate the limiting distances that would affect the stability of Dimorphos's interior structure or surface material. According to the classic theory of Roche (1847), a minimum distance exists for a purely fluid body below which tidal forces are greater than self-gravity, causing breakup. Holsapple & Michel (2006, 2008) provide a more specific theoretical framework, based on the Drucker–Prager strength model, suited to solid objects undergoing tidal stress, relying on parameters such as the internal friction angle and cohesion. Although these models provide theoretical insights to the tidal interaction problem, they both rely on assumptions that greatly simplify the treatment of the complex granular dynamics occurring within rubble-pile objects. In fact, it is reasonable to assume that rubble piles behave neither as a fluid nor as a solid, but rather as a complex granular system where long-range force chains may manifest and evolve. In this context, the theoretical predictions provided by Roche (1847) and Holsapple & Michel (2006, 2008) theories can be compared to the outcome of $N$-body granular simulations, which provide a better treatment of granular physics (e.g., Asphaug & Benz 1994; Movshovitz et al. 2012; Yu et al. 2014; Zhang & Michel 2020).

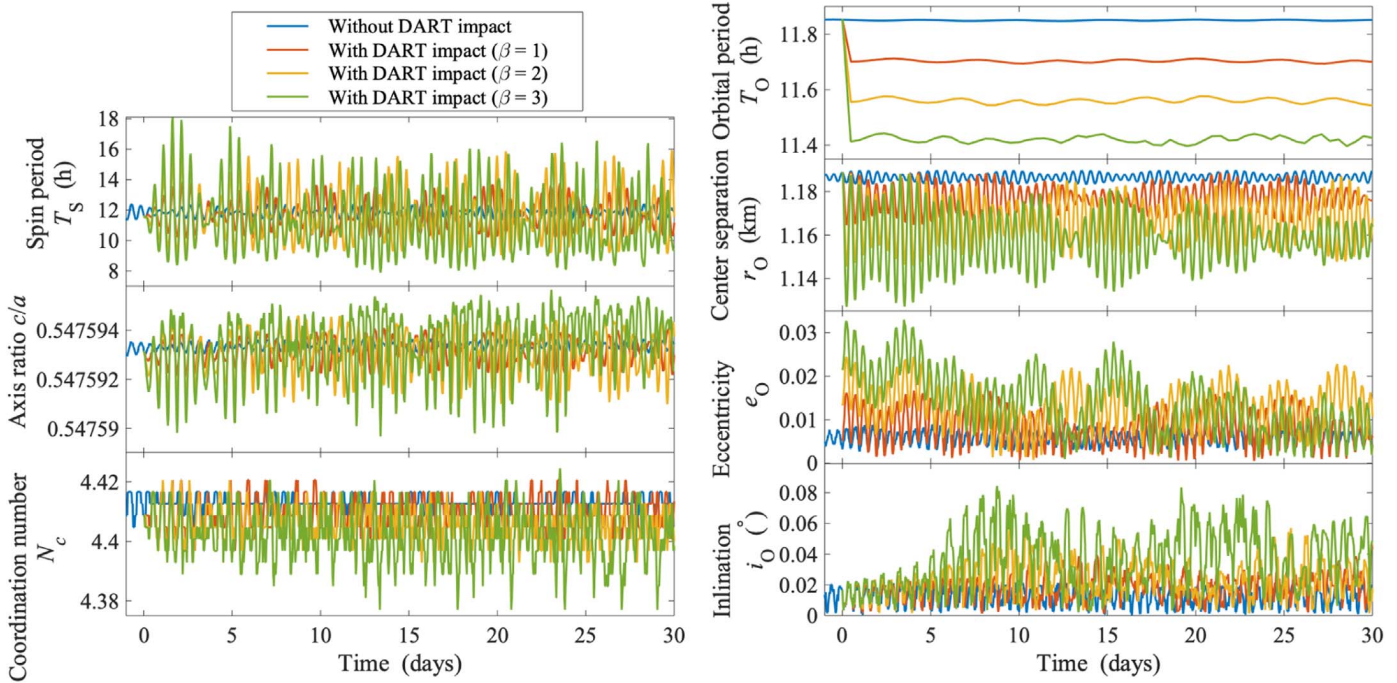

**Figure 9.** Evolution of Dimorphos's spin, shape, coordination number (i.e., the averaged contact number of the rubble pile), and orbital parameters of the full-rubble-pile model. The blue curves show the evolution without DART impact while the remaining three colors represent the results of instantaneous DART impacts with different $\beta$ values. We note that $e_O$ is the Keplerian eccentricity, computed at each output based on the instantaneous body position and velocity vectors.

In this section, we estimate the limiting distance between Dimorphos and Didymos where tidal effects are relevant to the stability of Dimorphos's internal structure and surface material. We use GRAINS to take advantage of the nonspherical shapes of the fragments, whose effect has shown to be relevant in short-term tidal interaction problems (e.g., the case of comet Shoemaker-Levy 9; Movshovitz et al. 2012). We model Dimorphos as a full rubble-pile object made of approximately 2000 meter-sized fragments. Tidal forces are automatically computed by the explicit $N$-body solver, as each fragment of rubble-pile Dimorphos interacts gravitationally with Didymos, which is modeled as a point-mass gravity source. We run 12 simulations in total, covering three different bulk density values (lower-bound 1820 kg m$^{-3}$; nominal 2170 kg m$^{-3}$; upper-bound 2520 kg m$^{-3}$) for Dimorphos. Each simulation lasts 12 hr (approximately one preimpact orbital period), and starts with Dimorphos at its nominal preimpact orbital location. The DART impact is modeled by means of an instantaneous change of Dimorphos's orbital velocity, which is selected to reach different pericenter distances (400, 500, 600, 800 m). For a direct comparison, we choose the pericenter distances to fall within the range provided by theoretical estimates based on Roche (1847); Holsapple & Michel (2006, 2008).

Figure 12 shows the qualitative results of GRAINS simulations and their direct comparison to theoretical estimates based on Roche (1847) and Holsapple & Michel (2008) models as a function of Dimorphos's bulk density. In particular, the Roche limit (represented by a green line) ranges between ∼600–700 m, depending on the bulk density of Dimorphos ($\rho_B = 2170 \pm 350$ kg m$^{-3}$). The colored region represents the disruption limits for a range of cohesion values, based on Holsapple & Michel (2006, 2008), and aggregates with a friction angle of 25°. In this case, the cohesionless breakup limit (upper edge of colored region) is consistently lower than the Roche limit for fluids. Also, according to Holsapple & Michel (2008) theory, the amount of cohesion needed to prevent breakup is extremely small, even at very low distances: e.g., a cohesion of 1 Pa is sufficient to prevent breakup at ∼400–450 m (blue line). The colored circles represent the outcome of GRAINS simulations in terms of their qualitative behavior on a four-level scale: *no effects*, where the aggregate's shape is largely preserved and little to no motion is observed on its surface; *reshaping*, where the aggregate's shape is considerably affected, but no mass loss is observed; *mass loss*, where some amount of mass (less than/equal to 50%) is lost by Dimorphos; and *disruption*, where more than 50% of

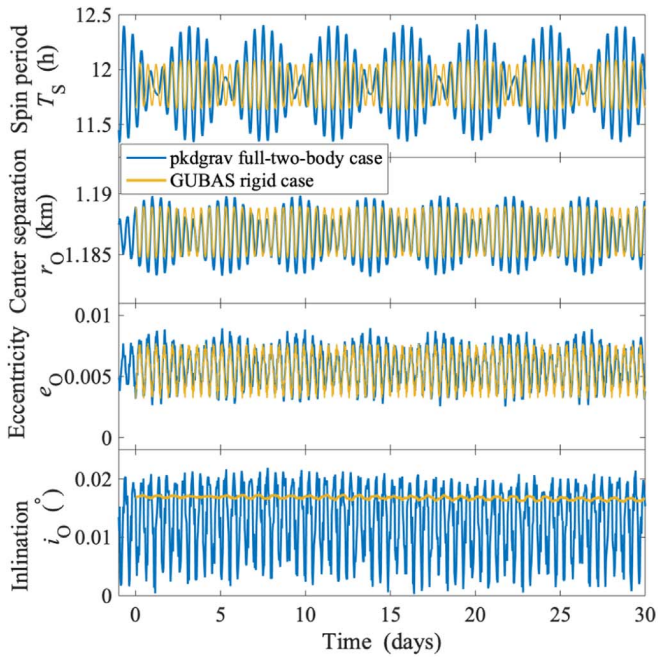
(a) Without DART impact (unperturbed).

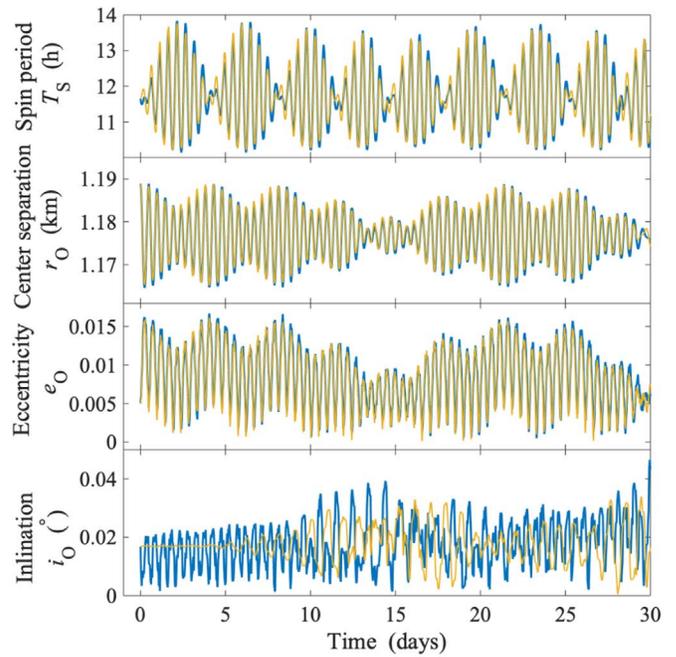
(b) With DART impact ($\beta = 1$).

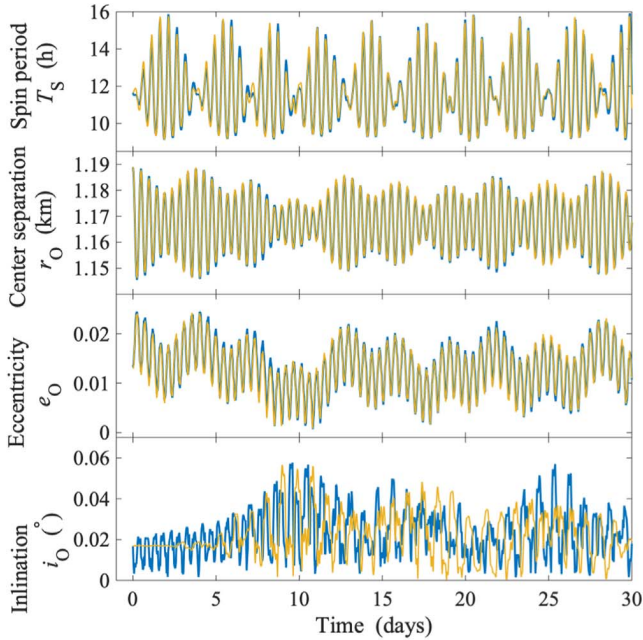
(c) With DART impact ($\beta = 2$).

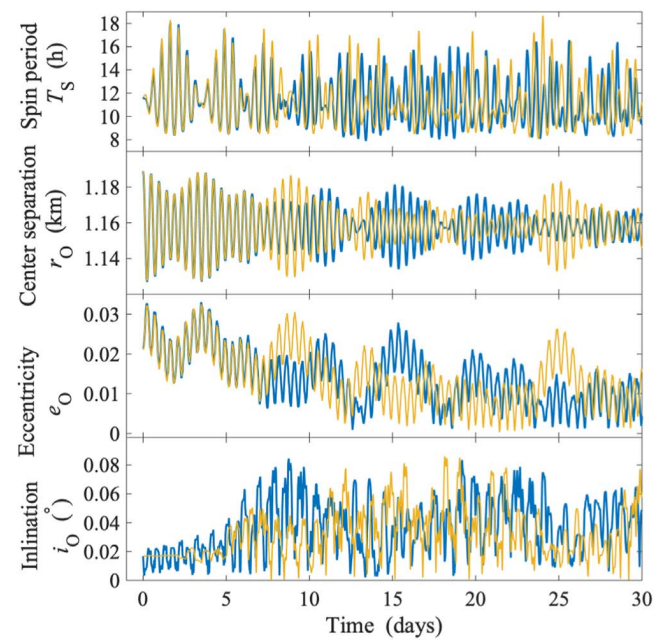
(d) With DART impact ($\beta = 3$).

**Figure 10.** Dimorphos's spin and orbital evolution for different $\beta$ values: the full-two-rubble-pile model (the blue curves) vs. GUBAS model (the yellow curves). The GUBAS and PKDGRAV simulations are in good agreement, both showing large orbital oscillation induced by the DART impact.

Dimorphos's mass is lost. We remark here that these results refer only to short-term tidal interactions, i.e., to effects observed within 12 hr from the simulated DART impact. There may be additional disturbances to Dimorphos's structure or surface on subsequent pericenter passages.

In this context, very weak or no effects are observed on aggregates orbiting with a pericenter of 800 m, while some effects are visible after a close passage at 600 m. In this case, depending on the density of Dimorphos, the aggregate experiences mass loss (lower-density case), or reshaping without mass loss (nominal- and higher-density cases). A pericenter at 500 m produces disruption of a low-density aggregate or a consistent mass loss in case the density is higher. For a lower pericenter, the consequences are more dramatic as the aggregate is completely shattered after a close passage at 400 m, for any density value within the range considered. Figure 13 shows the snapshots from three GRAINS simulations, with nominal bulk density of Dimorphos. They show the binary system 12 hr after instantaneous velocity change is applied, leading to, from left to right: a 600, 500, and 400 m pericenter orbit. As mentioned, the aggregate is heavily reshaped but with no mass loss in the 600 m case (left). On the other hand, we observe a significant mass loss in the 500 m case (center) and for lower pericenter distances, up to major shattering in the 400 m case (right). The physical extent of Didymos is shown: we recall that gravity is computed using a point-mass source, while the contact/collision interactions take into account the full polyhedral-shape model of Didymos, as shown in the figures. Thumbnails report the orbital trace of Dimorphos's barycenter in red, as well as the initial position of ellipsoidal Dimorphos in gray, before the instantaneous velocity change is applied.

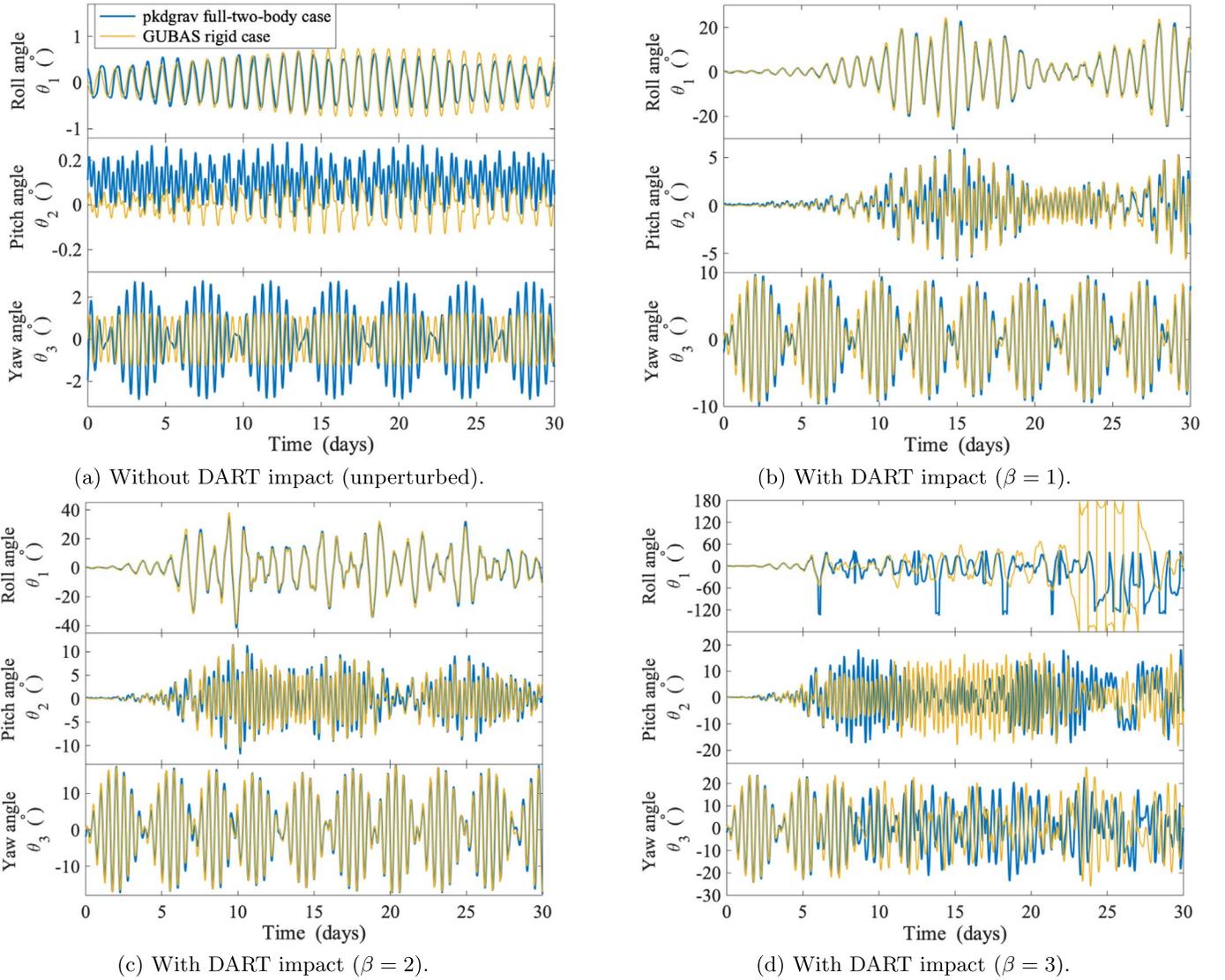

**Figure 11.** Evolution of Dimorphos's Euler angles with the full-two-rubble-pile model for different $\beta$ values. The GUBAS and PKDGRAV simulations are in good agreement, both showing large librations with about the same amplitudes and phases for $\beta = 1, 2$ and the barrel instability for $\beta = 3$.

Both the theoretical and numerical models predict strong tidal effects to occur in a close-proximity region, within 800 m from Didymos barycenter, i.e., approximately within a distance of 400 m from Didymos surface. As Dimorphos's preimpact orbit has a semimajor axis of approximately 1200 m, it is clear that a significant perturbation is required to reach the <800 m region. It is now worth interpreting these results in the context of the DART mission, by considering realistic orbit changes expected from DART impact itself. The $\beta$ value can be computed as a function of masses and speeds of both Dimorphos and DART:

$$\beta = \frac{\Delta v \cdot M_{\mathrm{B}}}{v_{\mathrm{DART}} \cdot M_{\mathrm{DART}}} \qquad (1)$$

where $\Delta v$ represents the change in Dimorphos's orbital speed produced by the DART impact. Using this simple relation, we compute $\beta$ values required to reach the region where strong tidal effects are relevant.

Table 3 reports the values of orbital period change ($\Delta T$), semimajor-axis change ($\Delta a$), and $\beta$ value required to reach a given pericenter distance (800, 600, 500, 400 m). It is worth noting that $\beta$ depends on Dimorphos's mass, which is highly uncertain and will only be measured with high accuracy by the Hera mission 4 yr after the impact. This uncertainty is propagated to the value of $\beta$, which varies by approximately 40% depending on Dimorphos's mass. We report $\beta$ values for the limiting cases of low ($\rho_{\mathrm{B,lo}} = 1820\,\mathrm{kg\,m^{-3}}$), nominal ($\rho_{\mathrm{B,nom}} = 2170\,\mathrm{kg\,m^{-3}}$), and high ($\rho_{\mathrm{B,hi}} = 2520\,\mathrm{kg\,m^{-3}}$) bulk density. As anticipated, the orbital change required to enter the region where strong tidal effects are relevant is very high. In the worst-case scenario (low Dimorphos mass), $\beta = 21$ is required to reach a 800 m close passage. This value is an order of magnitude higher than the typical expected values, where $\beta$ ranges typically from 1 to 5 and hardly reaches values as high as 10 (Stickle et al. 2022). Based on these considerations, it appears very unlikely for Dimorphos to enter the 800 m region, relevant for strong tidal effects, after DART's impact. Therefore, we can safely assume that both its internal structure and outer surface will not be affected in a relevant manner by short-term tidal interactions with Didymos, so long as $\rho_B$ is not significantly less than anticipated. However, we note that the simulations presented in this section are limited in their duration. Over longer integration times, it may be possible that Dimorphos's internal structure could be altered after subsequent pericenter passages as its spin and orbit evolves, even for the 800 m pericenter cases.

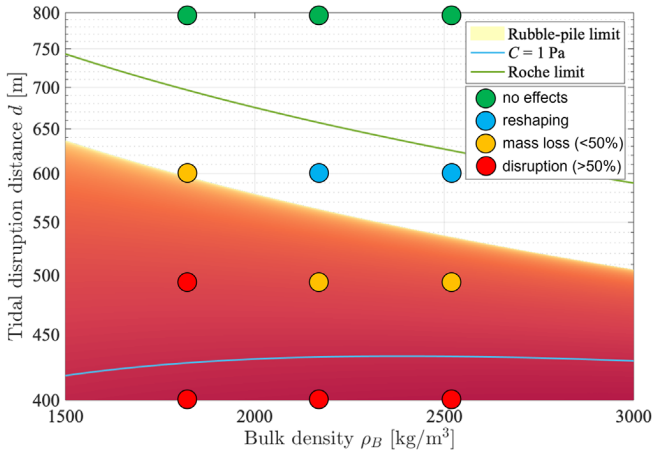

**Figure 12.** Tidal disruption distance as function of bulk density $\rho_B$ for the case of Dimorphos. The distance is considered here between the center of mass of Dimorphos and Didymos. The green line represents the Roche limit for fluid bodies (Roche 1847). The colored area represents disruption limits for aggregates with cohesion (the darker the color, the more cohesion is required to keep the aggregate stable) based on Holsapple & Michel (2006), Holsapple & Michel (2008), where the upper limit represents a cohesionless aggregate, and the blue curve shows an example for a cohesion of 1 Pa. Colored circles represent results of numerical simulations of cohesionless aggregates using GRAINS, where the colors are representative of the qualitative outcome of the simulation on a four-level scale (no effects/reshaping/mass loss/disruption).

### 4.2. Long-term Evolution of Dimorphos as an Irregularly Shaped Body

We also study the possibility of shape or surface changes to Dimorphos over longer timescales resulting from its post-impact rotation state. Previous rigid-body work has shown that the DART impact will cause Dimorphos to librate and possibly enter a chaotic rotation state, depending on its shape and $\beta$. Here we explore whether Dimorphos's post-impact rotation state can affect its internal structure or surface. We take a similar approach to the one described in Section 3.1, where we use PKDGRAV with a point-mass Didymos and rubble-pile Dimorphos. We also use the same particle-size—frequency distribution and friction parameters described in that section. However, instead of using the idealized ellipsoidal rubble piles, we instead use the radar-derived shape models of two near-Earth asteroids scaled to the volume of Dimorphos. Using a more realistic body shape allows parts of the surface to achieve higher surface slopes to allow us to find the reasonable limits at

**Table 3**
Orbital Variation Corresponding to a Post-impact Orbit of Dimorphos with Pericenter Values Ranging from 400 to 800 m

| Pericenter [m] | $\Delta T$ [hr] | $\Delta a$ [m] | $\beta$ $\rho_{B,lo}$ | $\rho_{B,nom}$ | $\rho_{B,hi}$ |
|---|---|---|---|---|---|
| 800 | 2.7 | 190 | 21 | 24 | 29 |
| 600 | 4.1 | 290 | 36 | 43 | 51 |
| 500 | 4.7 | 340 | 47 | 55 | 65 |
| 400 | 5.3 | 390 | 59 | 70 | 82 |

**Note.** $\beta$ values associated with such variations are computed for the limiting cases of low ($\rho_{B,lo} = 1820$ kg m$^{-3}$), nominal ($\rho_{B,nom} = 2170$ kg m$^{-3}$), and high ($\rho_{B,hi} = 2520$ kg m$^{-3}$) bulk density.

which the rubble-pile nature of the secondary may become important. The two radar-derived shape models we selected are Squannit, the secondary component of 66391 Moshup (Ostro et al. 2006), and 99942 Apophis (Brozović et al. 2018). Squannit was chosen because it is the secondary component of a system quite similar to Didymos and because its shape is expected to be attitude unstable in the Didymos system for an eccentricity of $e \sim 0.025$ ($\beta = 3$). A scaled Apophis-shape model was chosen because it has an irregular shape despite a relatively low shape elongation, and its DEEVE semiaxis lengths indicate that it should remain attitude stable within the Didymos system for $\beta = 3$.[9] PKDGRAV representations of these two bodies, scaled to the volume of Dimorphos, are shown in Figure 14 and Table 4 provides some quantitative descriptions of the two bodies.

Figure 15 is a mosaic of plots showing the evolution of the system for varying values of $\beta$ for the scaledApophis realization of Dimorphos. In each subfigure, the top plot shows Dimorphos's instantaneous spin rate along with the mutual orbital angular speed. The middle plot shows the instantaneous separation between Didymos and Dimorphos. The DART perturbation is applied after 24 hr of simulation time, which is why the orbital separation starts off near 1200 m, then drops to a lower value. The binary eccentricity is computed based on the periapse and apoapse of the orbit[10] (which can change due to spin–orbit coupling). The binary eccentricity is reported on each subfigure and is computed based on the first several orbit periods immediately following the DART perturbation. Finally, the third plot shows the change in Dimorphos's DEEVE semiaxis lengths, relative to their starting value. Rather than plot the changes to the body's moments of inertia, we instead plot the DEEVE axis lengths, since they have dimensions of length and are easier to conceptualize, although there is a straightforward correspondence between an arbitrarily shaped body and its DEEVE axis lengths.[11]

In Figure 15, we see that, in the scaledApophis case, the secondary's spin rate oscillates around the mean motion up to

---

[9] Assuming a uniform bulk density, Squannit's DEEVE axis lengths indicate it would be unstable within the Didymos system according to the analysis of Agrusa et al. (2021). Similarly, Apophis's DEEVE axis lengths indicate that it should be attitude stable as Didymos's secondary following a DART impact with $\beta = 3$.
[10] Since the mutual orbit is non-Keplerian, we report a geometric eccentricity rather than the Keplerian eccentricity. The geometric eccentricity of the orbit is a simple function of the maximum and minimum separations (periapse and apoapse) of the orbit. It can be written as $e = 1 - \frac{2}{r_a/r_p + 1}$, where $r_a$ and $r_p$ are the respective apoapse and periapse distances.
[11] For a body of mass $m$ and principal moments of inertia $A, B, C$, its corresponding dynamically equivalent equal-volume ellipsoid axis lengths $a, b, c$ are given by the following relations: $A = \frac{m}{5}(b^2 + c^2)$, $B = \frac{m}{5}(a^2 + c^2)$, $C = \frac{m}{5}(a^2 + b^2)$.

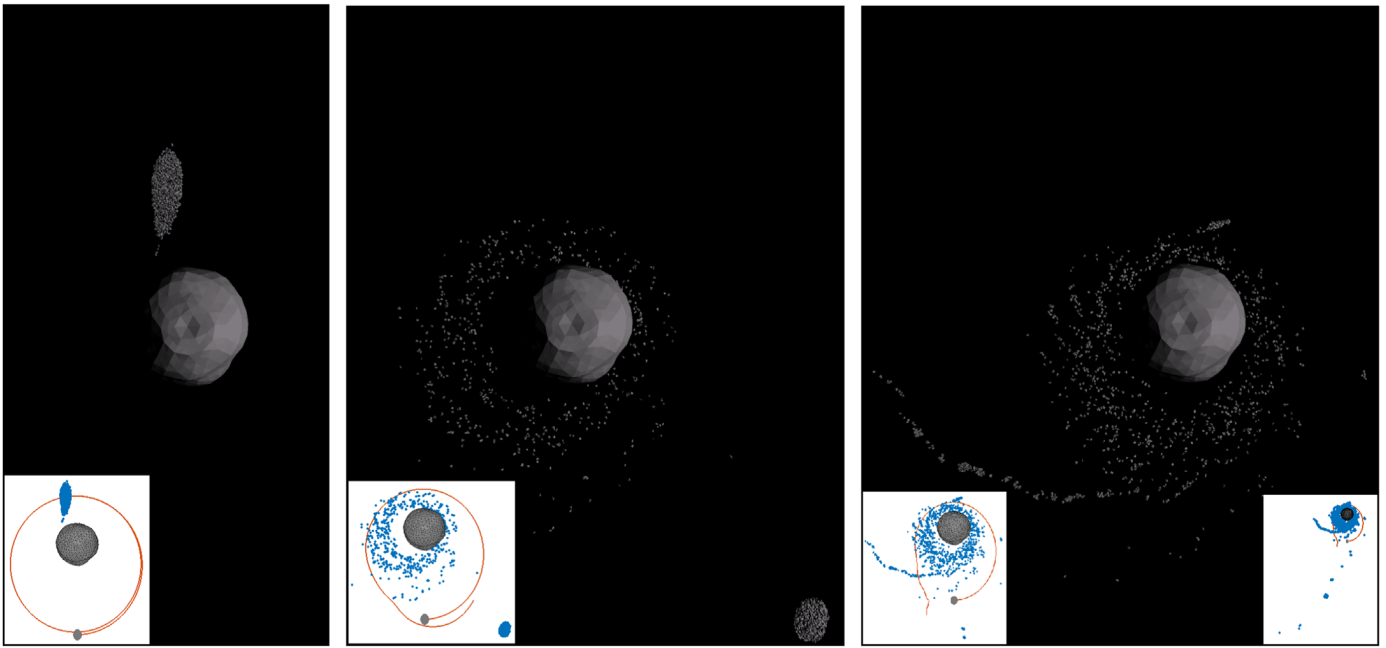

**Figure 13.** Snapshots of GRAINS simulations 12 hr after an instantaneous velocity change is applied on Dimorphos rubble-pile aggregate with nominal bulk density. From left to right: pericenter at 600 (reshaping with no mass loss), 500 (significant mass loss), 400 m (disruption). We show the physical extent of Didymos in scale for comparison. Didymos-shape model is used for collision computations, but not for gravity computations, which are done using a point-mass source. Thumbnails report the orbital trace of Dimorphos barycenter in red (including all particles that belonged to Dimorphos at the beginning of the simulation), and the initial position of Dimorphos before the instantaneous velocity change (gray ellipsoid).

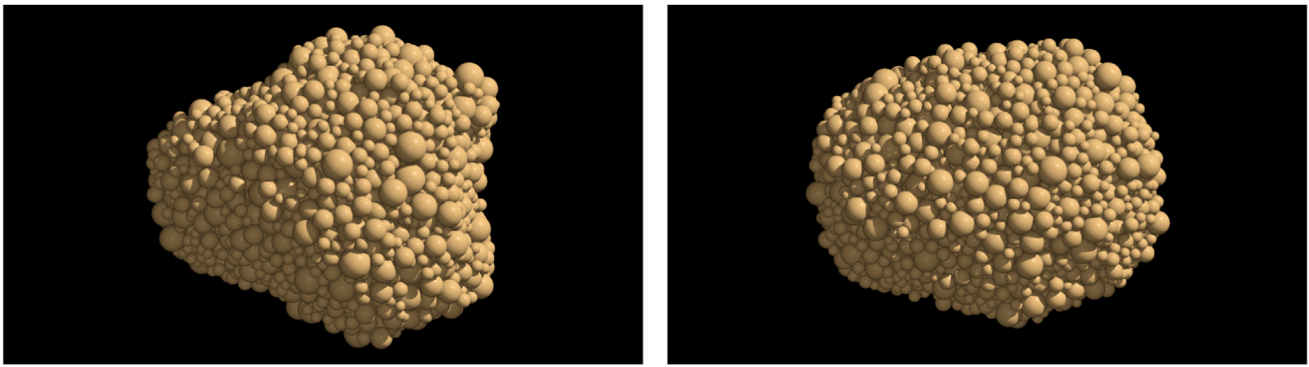

**Figure 14.** Top-down views of the rubble-pile models of Apophis and Squannit, scaled to the dimensions of Dimorphos. These bodies are built from the same particle-size distribution described in Section 3.1 and also have the same material parameters.

**Table 4**
The Body Shapes for Dimorphos Considered in This Section

| Name | $a/b$ | $b/c$ | $a$ [m] | $b$ [m] | $c$ [m] | $N$ | $M_B$ [$10^9$ kg] | Agrusa et al. (2021) Prediction for $\beta = 3$[a] |
|---|---|---|---|---|---|---|---|---|
| scaledApophis | 1.25 | 1.07 | 100.80 | 80.54 | 75.25 | 3801 | 4.852 44 | Stable |
| scaledSquannit | 1.32 | 1.32 | 109.68 | 82.97 | 63.02 | 3834 | 4.866 41 | Unstable |

**Note.** The shape models for the real Apophis (Brozović et al. 2018) and Squannit (Ostro et al. 2006) were scaled to match the expected volume of Dimorphos. The rubble-pile models for scaledApophis and scaledSquannit were simply created by deleting any PKDGRAV particles that lay outside the respective scaled shape models. The reported dimensions ($a$, $b$, $c$, $a/b$, $b/c$) are based on the DEEVE semiaxis lengths of the rubble piles and do not necessarily match the *exact* dimensions of the shape models used to create those rubble piles. $N$ and $M_B$ are the respective number of particles and mass of the body.
[a] These objects were not actually simulated in Agrusa et al. (2021), but their DEEVE semiaxes indicate stable/unstable attitudes respectively.

$\beta = 5$ since the body is stably librating. At $\beta = 7$, Dimorphos has become attitude unstable, and the spin rate can diverge from the mean motion. The same trend is true for the binary separation: the body separation uniformly oscillates due to the eccentricity of the orbit and the periodic exchange of angular momentum between Dimorphos's spin state and the mutual orbit. When Dimorphos becomes attitude unstable, its spin state becomes chaotic, as does the binary separation due to the strong spin–orbit coupling. The most interesting feature of these plots is the significant changes and oscillations in

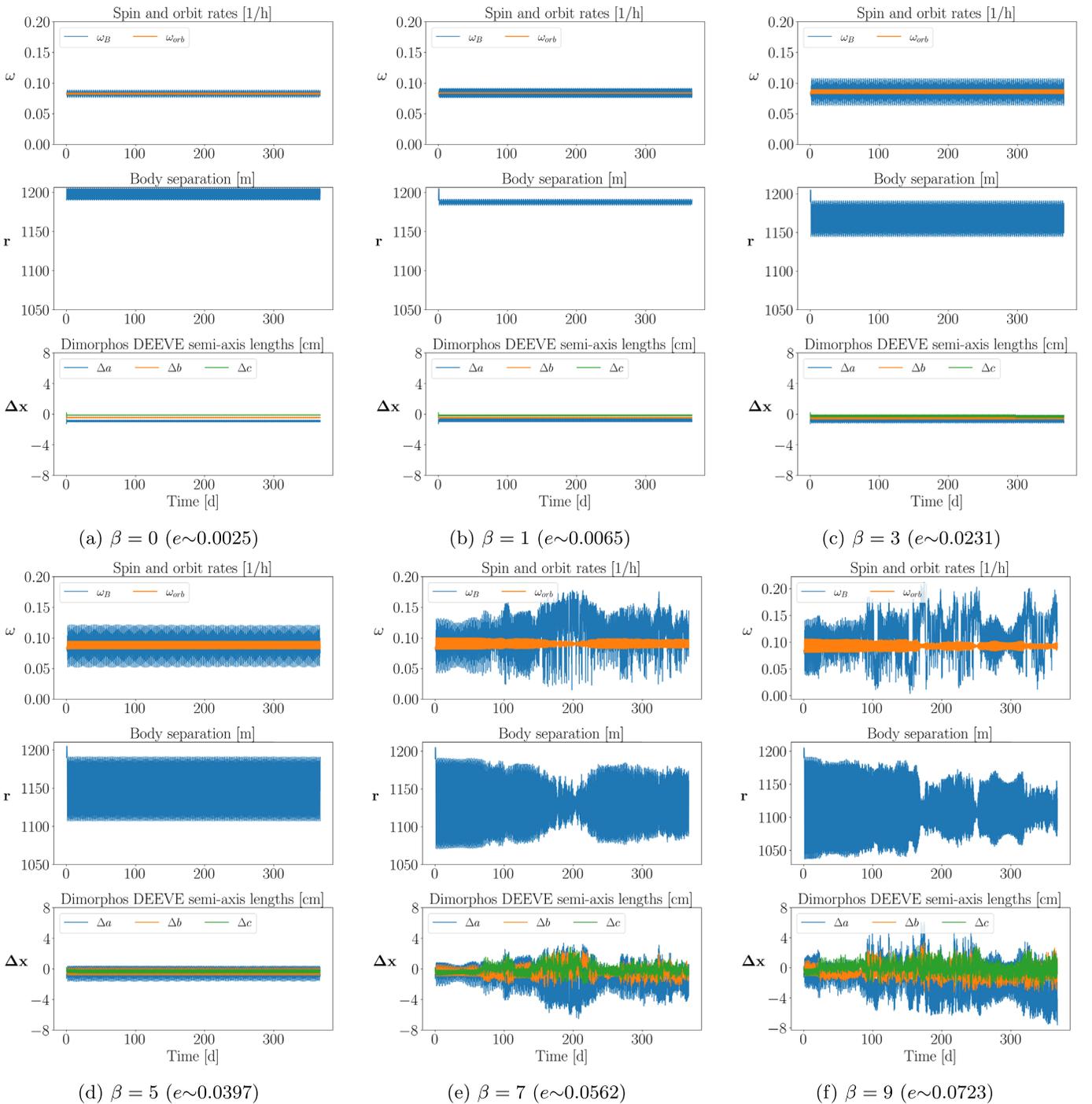

**Figure 15.** Evolution of the secondary's spin rate (top plots), orbital separation (middle plots), and change in DEEVE semiaxis lengths (bottom plots), for an Apophis-shaped Dimorphos with values of $\beta$ ranging from 0 to 9. For each value of $\beta$, we also report the binary orbital eccentricity $e$, based on the first several orbit periods following the DART-like perturbation. As $\beta$ increases, the body eventually becomes attitude unstable allowing relatively large deviations in the body's DEEVE axis lengths. All plots have the same y-axis scales to allow for direction comparisons.

Dimorphos's DEEVE axis lengths (i.e., moments of inertia). Even when Dimorphos is stably librating, the moments of inertia are also oscillating as the body feels time-varying stresses due to oscillations in the tidal potential and its spin state. This effect can be thought of as the body *breathing* or *flexing* in response to these stresses. This is a result of SSDEM, where particles are able to overlap by a small amount, which is mediated by a restoring spring force. As the tides or the body's spin change, the particles are able to make small adjustments to find a new equilibrium. When the body becomes attitude unstable, this effect becomes much more significant, and we see permanent changes to the DEEVE axis lengths indicating that some particles have actually been displaced, which will be discussed shortly.

Figure 16 shows the same mosaic of plots for the `scaledSquannit` realization of Dimorphos. Qualitatively, these plots are very similar, with the main difference being that Squannit becomes attitude unstable at lower values of $\beta$ due to its shape and thus has larger variations in its DEEVE semiaxis lengths.

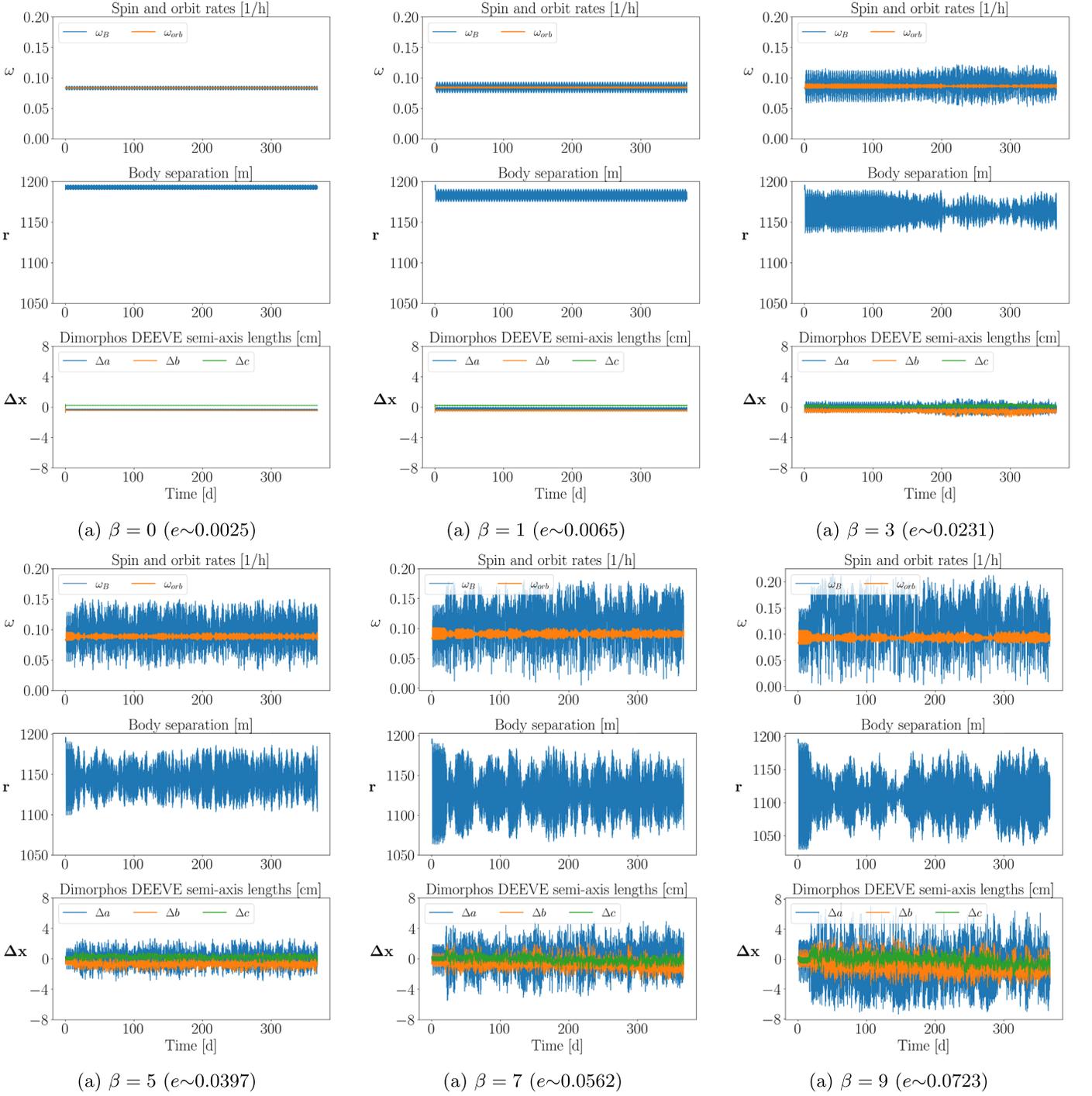

**Figure 16.** Evolution of the secondary's spin rate (top plots), orbital separation (middle plots), and change in DEEVE semiaxis lengths (bottom plots), for a Squannit-shaped Dimorphos with values of $\beta$ ranging from 0 to 9. For each value of $\beta$, we also report the binary orbital eccentricity $e$, based on the first several orbit periods following the DART-like perturbation. All plots have the same $y$-axis scales to allow for direction comparisons.

It should be noted that these results likely *exaggerate* the oscillations in the DEEVE axis lengths. In terms of axis-length change, the tidal response of a rubble pile is highly dependent on its Young's modulus, which relates its strain (axis-length change) to stress (applied force per unit area). In PKDGRAV, Young's modulus is not an input parameter but is related to the spring constant, $k_n$, that mediates particle overlaps. The Young's modulus ($Y$) can be approximated as $Y \sim \frac{k_n}{\pi R}$, where $R$ is the typical particle radius (DeMartini et al. 2019). In other words, a weaker spring constant means the material is weaker and easier to deform. In these simulations, the average particle radius is 4.2 m, and the spring constant is $k_n \sim 1.45 \times 10^4$ N m$^{-1}$, corresponding to a Young's modulus of $\sim$1.1 kPa. This value is quite small for granular material, although not completely unrealistic; a value on the order of 1 MPa or greater is probably more realistic (Möhlmann et al. 2018). The low value for the spring constant (and therefore Young's modulus) was chosen out of computational necessity.

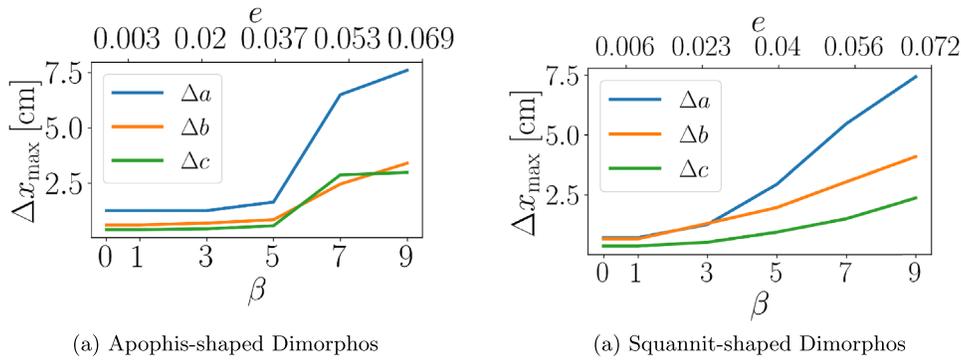

**Figure 17.** Maximum DEEVE axis-length changes as a function of $\beta$ or the binary eccentricity $e$. The axis-length change for $\beta = 0$ is nonzero since the system is not *perfectly* relaxed and has a nonzero eccentricity and libration. As $\beta$ or $e$ is increased, the DEEVE axis lengths have larger oscillations, and the trend jumps up sharply when the body becomes attitude unstable. In these simulations, Dimorphos's Young's modulus is ∼1 kPa and likely *overestimates* the axis-length changes for reasons explained in the text. For higher values of the Young's modulus, we would expect lower magnitudes in axis-length oscillations, although the general trend would not change.

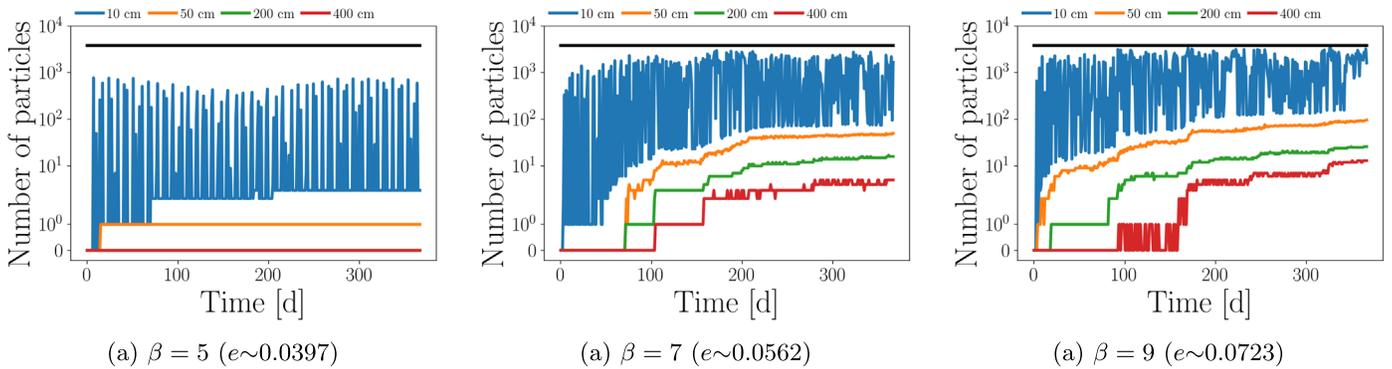

**Figure 18.** Particles that have moved by more than a given distance for the scaled Apophis case. As $\beta$ (or equivalently the binary eccentricity $e$) is increased, more particles are able to move from their original location. These motions are ongoing and occur hundreds of orbit periods after the DART perturbation is applied. However, the average particle radius is 420 cm, meaning that these motions are very small. The large number of particles that are moving by 10 cm or more indicates that we are not sensitive to motion at this scale due to the combination of limited particle resolution and the fact that the body's moments of inertia (which form the fixed-body coordinate frame) are changing.

A higher spring constant means higher restoring forces between particles, which requires a shorter time step to adequately resolve the particle interactions (Schwartz et al. 2012). This means that the time step needs to be reduced by a factor of 10 if we want to increase the spring constant (and therefore the Young's modulus) by a factor of 10, which makes long-term simulations too slow.[12] Conveniently, previous work has shown that the axis-length changes scale linearly with the Young's modulus (DeMartini et al. 2019). So if Dimorphos has a Young's modulus of $Y \sim 1$ MPa, then we would expect the deviations in its DEEVE axis lengths to be a factor of $\sim 10^{-3}$ times smaller than what is shown here, for example. Figure 17 shows the maximum change in axis length over the entire simulation as a function of $\beta$ for the two body shapes under consideration. Although the artificially low Young's modulus exaggerates the magnitudes in the axis length change, these plots illustrate the strong dependence on $\beta$. An equivalent plot with a more realistic Young's modulus would qualitatively look similar, with the y-axis scaled to lower values.

Based on these simulations, it seems that the oscillations in Dimorphos's DEEVE axis lengths are caused by time-varying stresses due to spin and tides, while the magnitude of the oscillations is highly dependent on the material properties of the body. Now, we turn to briefly investigate the cause of the permanent changes to the DEEVE axis lengths. We find that the cause is likely small particle motions on the surface of the body that lead to small changes in the body's moments of inertia (and therefore DEEVE semiaxis lengths). In Figure 18, we plot the number of particles that have moved by a given distance for the Apophis-shaped Dimorphos, for $\beta = 5, 7, 9$. In order to calculate whether a particle has moved, we compute its position in the body-fixed frame at each output and compare it to its position when the DART perturbation was first applied. It is important to keep in mind that the body-fixed frame is constructed by computing the body's principal rotation axes at each time step, which depend on the body's moments of inertia. Because the principal axes are able to change in direction, the body-fixed frame is not perfectly fixed! Due to this effect, this approach is not sensitive to small particle motions. However, the average particle size is 4.2 m, meaning that the sensitivity to small-scale motion is limited anyways.

In Figure 18(a) ($\beta = 5$), we see that no particles have moved more than 400 or 200 cm, one has moved more than 50 cm, and many particles have moved by 10 cm or more. Figures 18(b) and 18(c) show that more particles are able to move from their original location when the binary eccentricity (i.e., $\beta$) is increased. The same results are shown in Figure 19 for the Squannit-shaped Dimorphos with largely the same conclusion:

---

[12] The simulations presented here used 4 CPU cores in parallel, and took ∼3 months to complete a 1 yr simulation. All things being equal, it would take ∼250 yr of wall-clock time to repeat these simulations with a more realistic Young's modulus of ∼1 MPa.

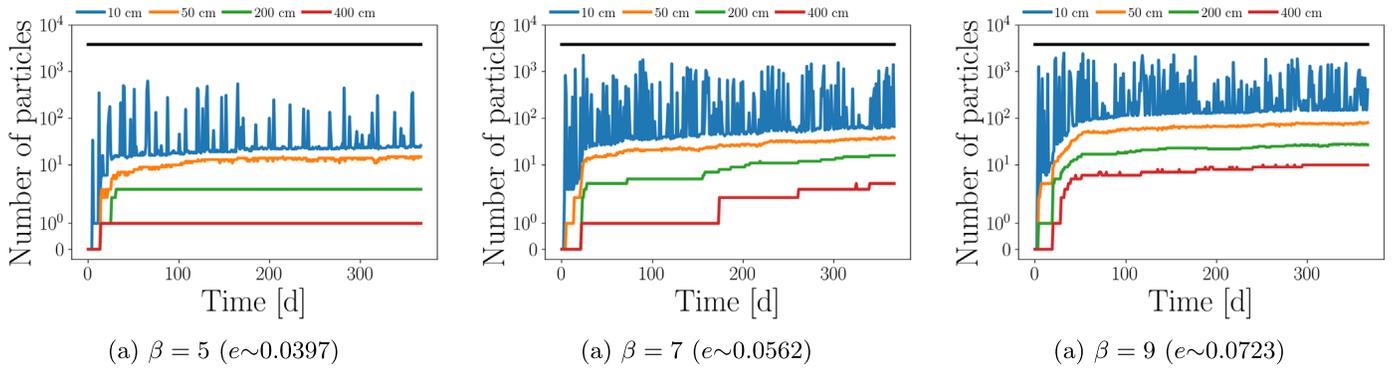

Figure 19. Particles that have moved by more than a given distance for the scaled Squannit case.

a more eccentric and tighter orbit leads to more particle motion. It is important to keep in mind that the typical particle radius is 4.2 m, meaning that these particles are not traveling large distances on the surface. Rather, these are particles making small adjustments, typically moving less than a single particle radius.

*4.2.1. Interpretation of Results*

Here, we provide a guide for interpreting the results of this subsection in the context of the various choices and compromises in simulation settings.

1. The large oscillations in the number of particles that move ⩾ 10 cm (Figures 18 and 19) indicate that the simulations are not sensitive to small-scale particle motion. This is due to a combination of the coarse particle resolution and the fact that Dimorphos's body-fixed frame is not technically fixed. At each simulation output, Dimorphos's body-fixed frame is defined using its principal rotation axes, which depend on its time-varying inertia tensor.
2. The artificially low value of the Young's modulus (∼1 kPa here) means that the simulated rubble piles are likely overly deformable, meaning that we are *overestimating* the tidal response of Dimorphos. This choice in Young's modulus was required to reduce the computational costs. However, it has been shown that Young's modulus is directly proportional to the maximum axis-length change (DeMartini et al. 2019). This means that the DEEVE axis-length changes can be scaled down to approximate a more realistic axis-length change. Something of the order of millimeters (or less) seems probable. Such a small effect is unlikely to be measurable with Hera. Though, in theory, an in situ seismometer would be capable of detecting this signal.
3. The coarse particle size (∼4 m radii) was chosen to limit the computational costs as well. This means these simulations likely *underestimate* the particle motion on the surface. This is because many surface particles are sitting in deep gravitational wells and have to be lifted out of a meters-deep crevice before they can move along the surface. Of course, if DART imagery indicates a lack of sub-meter-sized boulders, then this is no longer an issue as it would represent a realistic treatment of Dimorphos's surface. However, if Dimorphos has smaller-sized particles on its surface, then a higher-resolution simulation (or an alternative approach) is necessary to fully quantify the amount of surface motion that may occur as a result of Dimorphos's post-impact dynamical state.
4. All these simulations use *gravel-like* friction parameters that yield a friction angle of $\phi \sim 38°$—a lower friction angle would increase the odds that a portion of Dimorphos's surface could exceed its angle of repose while tumbling, leading to surface motion.

For these reasons, we conclude that DART-induced particle motion on Dimorphos is a possibility under the right circumstances, although a more detailed investigation is required. In addition, the fact that the tumbling, and therefore particle motion, is long-lived indicates that ESA's Hera mission may observe this effect in real time upon rendezvous in 2026. After DART's arrival, the parameter space of possible body shapes and boulder-size ranges will shrink dramatically (Daly et al. 2022), making this problem more tractable. If preliminary DART imagery suggests that Dimorphos is indeed a rubble pile, that its body shape indicates post-impact tumbling, and that it has a shape with high surface slopes, then the possibility of surface particle motion will be explored with much higher fidelity. Full-scale PKDGRAV-like simulations with a more realistic Young's modulus and particle-size distribution, in addition to high-resolution localized granular bed simulations at particular locations on Dimorphos's surface, could be used to investigate the likelihood of long-term surface particle motion. However, given that Dimorphos's shape and boulder-size–frequency distribution are still unknown, such a high-fidelity study is unwarranted at this point.

## 5. Conclusions and Future Work

In Section 3.1, we conducted long-term (1 yr) simulations with Dimorphos as a rubble pile having the shape of a triaxial ellipsoid following the DART impact. We found that, for a momentum enhancement factor, $\beta$, on the order of 3, Dimorphos's evolution as a rubble pile is not appreciably different than its evolution as a rigid body. This holds true in cases where Dimorphos is in a stable libration state and when it rotates chaotically. We extended this study in Section 3.2 to include a rubble-pile treatment for Didymos and found no substantial differences, although these simulations were limited to only 30 days due to the increased number of particles and computational cost. These results indicate that the much faster rigid-body approach is an appropriate tool for modeling the post-impact dynamics of the Didymos binary following the DART impact.

In Section 4, we explored the limits at which a rubble-pile treatment might be necessary. In Section 4.1, we showed that $\beta$ would have to be unrealistically large ($\beta \gtrsim 20$) in order for the mutual tides to cause significant structural changes to Dimorphos, at least over short timescales and for the expected range in bulk density. We then simulated Dimorphos as an irregular shape using scaled shape models of Squannit and Apophis over a wider range of $\beta$ values ($0 < \beta < 9$) to place rough constraints on how the excitation due to DART will affect Dimorphos over longer timescales. We found that the mass distribution of Dimorphos (i.e., its moments of inertia), measurably changes in response to the time-varying spin and tidal environment. Even for small values of $\beta$, we observed the DEEVE semiaxis lengths of Dimorphos oscillate as its spin rate and orbital separation are periodically changing. However, the magnitude of these oscillations are probably overexaggerated in these simulations due to the unavoidable selection of material parameters. Oscillations on the order of millimeters are probably more likely than the centimeter-scale oscillations presented here, although this is all highly dependent on Dimorphos's (unknown) material strength and interior structure. For $\beta \gtrsim 5$, we also saw small permanent changes to the semiaxis lengths as a result of small motions of particles at the surface. Due to Dimorphos's rapidly changing spin state when its attitude becomes chaotic, as well as the periodically varying tidal force, these particles feel high-enough accelerations that enable them to move on the surface. These motions are small and are typically less than a single particle radius. However, the limited particle resolution underestimates this effect, making it difficult to draw any firm conclusions. In any case, these simulations indicate that, under the right circumstances, the motion of particles on the surface is plausible and that a more focused investigation is required. If surface motion does occur, we showed that it could be a long-lived process, meaning that it is something that Hera may be able to observe upon arrival in late 2026. There is also the prospect that any regolith motion could alter the shape of the DART impact crater, prior to Hera's arrival in a manner analogous to surface refreshment at Stickney Crater on Phobos (Ballouz et al. 2019).

To summarize, we identify four key results of this work:

1. We find that a rubble-pile approach is not required to capture Dimorphos's post-impact spin and orbital evolution, so long as $\beta$ is not unexpectedly large. Therefore, faster rigid-body codes should be more than adequate for predicting the system's post-impact dynamical state.
2. If $\beta$ is *significantly* larger than expected ($\beta \gtrsim 20$) or Dimorphos is highly underdense (or undermassive), then significant reshaping, mass loss, or disruption would be possible. However, we find that such an outcome is unlikely given experimental and numerical predictions for $\beta$ (Stickle et al. 2022; Walker et al. 2022).
3. Depending on the internal structure and material properties of Dimorphos and the level of excitation to its spin and orbital state, a large tidal response may be induced. In theory, this could be measured with a seismometer. In practice, this effect will likely not be measurable with Hera.
4. Depending on Dimorphos's shape and post-impact spin state, granular motion on the surface is a possibility. We expect that the methodology used in this work *underestimates* any motion on the surface. This effect requires more thorough investigation, which is currently underway. If surface motion occurs, it presents a unique possibility that Hera may observe this effect by characterizing the surface color and grain flow patterns.

This work was supported in part by the DART mission, NASA Contract #80MSFC20D0004 to JHU/APL. F.F. acknowledges funding from the Swiss National Science Foundation (SNSF) Ambizione grant No. 193346, and the National Centre for Competence in Research PlanetS supported by the Swiss National Science Foundation. P.M. acknowledges funding support from the European Union's Horizon 2020 research and innovation program under grant agreement No. 870377 (project NEO-MAPP), from the French space agency CNES and the European Space Agency (ESA).

We kindly thank the reviewers, whose comments significantly improved the manuscript. Some of the simulations herein were carried out on The University of Maryland Astronomy Departments YORP cluster, administered by the Center for Theory and Computation. Raytracing for Figures 2, 8, and 14 were performed using the Persistence of Vision Raytracer.

## ORCID iDs

Harrison F. Agrusa  https://orcid.org/0000-0002-3544-298X
Fabio Ferrari  https://orcid.org/0000-0001-7537-4996
Yun Zhang  https://orcid.org/0000-0003-4045-9046
Derek C. Richardson  https://orcid.org/0000-0002-0054-6850
Patrick Michel  https://orcid.org/0000-0002-0884-1993

## References

Agrusa, H. F., Gkolias, I., Tsiganis, K., et al. 2021, Icar, 370, 114624
Agrusa, H. F., Richardson, D. C., Davis, A. B., et al. 2020, Icar, 349, 113849
Asphaug, E., & Benz, W. 1994, Natur, 370, 120
Ballouz, R.-L., Baresi, N., Crites, S. T., Kawakatsu, Y., & Fujimoto, M. 2019, NatGe, 12, 229
Brozović, M., Benner, L. A. M., McMichael, J. G., et al. 2018, Icar, 300, 115
Cheng, A. F., Raducan, S. D., Fahnestock, E. G., et al. 2022, PSJ, 3, 131
Cheng, A. F., Rivkin, A. S., Michel, P., et al. 2018, P&SS, 157, 104
Ćuk, M., & Burns, J. A. 2005, Icar, 176, 418
Ćuk, M., Jacobson, S. A., & Walsh, K. J. 2021, PSJ, 2, 231
Daly, R. T., Ernst, C. M., Barnouin, O. S., et al. 2022, PSJ, in press
Davis, A. B., & Scheeres, D. J. 2020, Icar, 341, 113439
Davis, A. B., & Scheeres, D. J. 2021, GUBAS: General Use Binary Asteroid Simulator, Astrophysics Source Code Library, ascl:2107.013
DeMartini, J. V., Richardson, D. C., Barnouin, O. S., et al. 2019, Icar, 328, 93
Dotto, E., Della Corte, V., Amoroso, M., et al. 2021, P&SS, 199, 105185
Fang, J., & Margot, J.-L. 2012, AJ, 143, 24
Ferrari, F., Lavagna, M., & Blazquez, E. 2020, MNRAS, 492, 749
Ferrari, F., & Tanga, P. 2020, Icar, 350, 113871
Ferrari, F., & Tanga, P. 2022, Icar, 378, 114914
Ferrari, F., Tasora, A., Masarati, P., & Lavagna, M. 2017, Multibody Syst. Dyn., 39, 3
Fleischmann, J., Serban, R., Negrut, D., & Jayakumar, P. 2015, J. Comput. Nonlinear Dyn., 11, 044502
Fuentes-Muñoz, O., & Scheeres, D. J. 2020, in AIAA Scitech 2020 Forum (Reston, VA: AIAA)
Goldreich, P., & Sari, R. 2009, ApJ, 691, 54
Hirabayashi, T., Ferrari, F., Jutzi, M., et al. 2022, PSJ, 3, 140
Holsapple, K. A., & Michel, P. 2006, Icar, 183, 331
Holsapple, K. A., & Michel, P. 2008, Icar, 193, 283
Jacobson, S. A., & Scheeres, D. J. 2011, Icar, 214, 161
Meyer, A. J., Gkolias, I., Gaitanas, M., et al. 2021, PSJ, 2, 242
Meyer, A. J., & Scheeres, D. J. 2021, Icar, 367, 114554
Michel, P., Kueppers, M., Sierks, H., et al. 2018, AdSpR, 62, 2261
Michel, P., Küppers, M., Bagatin, A. C., et al. 2022, PSJ, in press
Michikami, T., Honda, C., Miyamoto, H., et al. 2019, Icar, 331, 179
Michikami, T., Nakamura, A. M., & Hirata, N. 2010, Icar, 207, 277

Möhlmann, D., Seidensticker, K. J., Fischer, H.-H., et al. 2018, Icar, 303, 251
Movshovitz, N., Asphaug, E., & Korycansky, D. 2012, ApJ, 759, 93
Naidu, S. P., Benner, L. A. M., Brozovic, M., et al. 2020, Icar, 348, 113777
Naidu, S. P., Chesley, S. R., Farnocchia, D., et al. 2022, PSJ, submitted
Nakano, R., Hirabayashi, M., Agrusa, H. F., et al. 2022, PSJ, 3, 148
Nimmo, F., & Matsuyama, I. 2019, Icar, 321, 715
Ostro, S. J., Margot, J.-L., Benner, L. A. M., et al. 2006, Sci, 314, 1276
Pajola, M., Barnouin, O., Lucchetti, A., et al. 2022, PSJ, submitted
Pazouki, A., Kwarta, M., Williams, K., et al. 2017, PhRvE, 96, 042905
Pravec, P., Scheirich, P., Kušnirák, P., et al. 2016, Icar, 267, 267
Pravec, P., Thomas, C. A., Rivkin, A. S., et al. 2022, PSJ, in press
Quillen, A. C., LaBarca, A., & Chen, Y. 2022, Icar, 374, 114826
Raducan, S. D., & Jutzi, M. 2022, PSJ, 3, 128
Richardson, D. C., Quinn, T., Stadel, J., & Lake, G. 2000, Icar, 143, 45
Richardson, D. C., Agrusa, H. F., Barbee, B., et al. 2022, PSJ, 3, 157
Rivkin, A. S., Chabot, N. L., Stickle, A. M., et al. 2021, PSJ, 2, 173
Roche, E. 1847, Acad. Sci. Lett. Montpelier. Mem. Section Sci., 1, 243
Scheirich, P., & Pravec, P. 2009, Icar, 200, 531
Scheirich, P., & Pravec, P. 2022, PSJ, in press
Schwartz, S. R., Richardson, D. C., & Michel, P. 2012, Granular Matter, 14, 363
Stadel, J. G. 2001, PhD thesis, Univ. Washington
Stickle, A. M., Burger, C., Caldwell, W. K., et al. 2022, PSJ, submitted
Tasora, A., Serban, R., Mazhar, H., et al. 2016, in High Performance Computing in Science and Engineering. HPCSE 2015. Lecture Notes in Computer Science, Vol. 9611, ed. T. Kozubek et al. (Cham: Springer), 19
Walker, J. D., Chocron, S., Grosch, D. J., et al. 2022, PSJ, submitted
Walsh, K. J. 2018, ARA&A, 56, 593
Walsh, K. J., Jawin, E. R., Ballouz, R. L., et al. 2019, NatGe, 12, 242
Walsh, K. J., Richardson, D. C., & Michel, P. 2008, Natur, 454, 188
Wisdom, J. 1987, AJ, 94, 1350
Yu, Y., Richardson, D. C., Michel, P., Schwartz, S. R., & Ballouz, R.-L. 2014, Icar, 242, 82
Zhang, Y., & Michel, P. 2020, A&A, 640, A102
Zhang, Y., Michel, P., Richardson, D. C., et al. 2021, Icar, 362, 114433
Zhang, Y., Richardson, D. C., Barnouin, O. S., et al. 2017, Icar, 294, 98
Zhang, Y., Richardson, D. C., Barnouin, O. S., et al. 2018, ApJ, 857, 15